\documentclass[%
reprint,
 amsmath,amssymb,
 aps, physrev,
prstper,
floatfix,
]{revtex4-2}

\usepackage{graphicx}
\usepackage{dcolumn}
\usepackage{bm}

\usepackage{xcolor}

\begin{document}

\title{\textbf{Missing Data on Physics Exams: Demographic Patterns, Course-Level Predictions, and Implications for Equity}
}%

\author{Cassandra Paul}
\affiliation{Department of Physics \& Astronomy, Science Education Program, Undergraduate Advising and Success, San Jose State University, One Washington Square, San Jose, CA, 95192}

\author{David J. Webb}
\affiliation{Department of Physics \& Astronomy, University of California, Davis, One Shields Avenue, Davis, CA, 95616}
\author{Mary Kate Chessey}
\affiliation{Barcelona Supercomputing Center, Barcelona, Spain} 

\date{\today}

\begin{abstract}
In a previous quantitative retrospective study we showed that different demographic groups of students leave different numbers of problems blank on physics exams, leading to inequities in course outcomes. In that work we argued that there were good reasons to treat these blanks as missing data, rather than indicators of a lack of understanding. In this paper, we refine this analysis and show more detailed breakdowns of uncollected test item responses by race/ethnicity and first generation college student status, coming to the same conclusion: test item responses are uncollected for students with different ethnic and racial backgrounds at different rates, and these patterns are not exclusive to low-performing students. We also correct an error from our previous work, finding here that there is no significant gender difference in uncollected test item responses. Finally, we provide a more robust analysis of course level data illustrating that blanks are a variable controlled at the course level rather than the student level, providing more evidence for the use of a course deficit model (rather than a student deficit model) when examining equity disparities, and also suggesting that there are plausible means for instructors to minimize uncollected test item responses, and therefore reduce or even eliminate the bias associated with this missing data. We provide a couple suggestions for faculty who want to minimize the impacts of blanks.

\end{abstract}

\maketitle


\section{\label{sec:Intro}Introduction}
Grades are used by students, teachers, schools, and much of the rest of society as a measure of students' understanding of, or skills in, a subject. Given this expectation, it’s useful to understand the connection between a grade given to a student and the actual understanding or skill of that student. This connection is made more difficult when an assessment fails to capture information about a person's understanding for any reason, for example, when open-ended test items are left blank, and therefore are not collected for evaluation.

There is little research on blank responses on open ended problems. Matters et al. \cite{mattersMultipleChoiceShortResponseItems1999} find that students are much more likely to omit open ended responses than multiple choice questions, even when the open ended problems are worth much more than the multiple choice questions. Chiacchio et al. \cite{dichiacchioExaminingHowMotivation2016} distinguish between patterns of uncollected test item data by identifying ``leavers'', test-takers who disengage from completing items after a certain point, and ``jumpers'', those who skip some problems throughout the exam, suggesting that students might leave blanks for different reasons. What research does exist, is outside the field of PER, with the exception of the authors' own conference paper \cite{Paul2018} from 2018.

In our previous analysis of the data-set we present in this paper, we discussed ``blank-leaving behavior'' associated with the student \cite{Paul2018}. In this paper we adjust our language and instead position the blank test items as a failure of the exam to collect meaningful assessment information from the student. This shift is purposeful in the sense that it focuses the conversation around changing assessments and grading rather than faulting the students. This change is also informed by new results shared in this paper indicating that the rates of uncollected test item responses (or the rates of blank responses) vary tremendously by course section, implying that assessments themselves play a big role in the rate of uncompleted items.



Quizzes and exams may fail to measure a person's understanding for extraneous reasons, for example, when insufficient time is allocated for completion of the exam, test items are unclear or written in an ambiguous manner, printing or formatting obscured the presence of some test items, or for any other number of reasons. Often, test-takers submit test forms with uncompleted items.


One way instructors commonly deal with a blank test item is grading the item as a zero, likely reasoning that if the student doesn’t provide any evidence of understanding, no credit can be awarded. While on the surface this instructor practice seems logical and fair, it is not a measurement of the student's knowledge of a given topic. As instructors, the authors of this paper have often wished that students had written something down so that there was some data to evaluate the students' responses. However, because of the nature of the format of a written exam, a conversation is not possible at this point. Thus a zero, the numerical equivalent of no understanding, is awarded for what instead might be better considered missing data. 

An instructor might be content with awarding a grade that includes both student behavior and understanding (in fact the authors of this paper have done this as instructors) however, if test item responses for different student populations are collected at different rates, and this practice negatively impacts some student groups more than others, then assigning zeros to uncollected item responses may well contribute to inequity in the course. The focus of this work is on the differential prevalence of uncollected test item responses by student demographic group and their impact on overall course grades.




\section{\label{sec:Inequitable Course Structures }Inequitable Course Structures}

While decades of equity research focuses on demographic differences in opportunities and preparation being the cause of equity gaps, there is a growing body of research indicating that an additional major contributor to equity gaps are course structures themselves \cite{Cotner2017, decaro_active_2025, webb_attributing_2023, paul_examining_2025, young_reducing_2025}. In other words, it is not that courses are just measuring differences in performance that are the result of past inequities, instead, courses may in fact be creating inequities in the way that they are choosing to present material and/or make measurements. 


To understand how or why measurement choices can cause equity gaps, it's helpful to consider that different demographic groups have different kinds of experiences outside of school (such as values, and cultural practices) and different assessment practices can be more or less aligned with those experiences. Various studies \cite{stephens_unseen_2012,stathopoulou_language_2006} over the years have investigated the extent to which a mismatch between the norms of a demographic group of students and the norms of their educational institution may affect learning and the resulting grades of that group of students. For example, Stephens et al. \cite{stephens_unseen_2012} suggest that first-generation college students are perhaps oriented more toward values necessary for a community and less toward the individual independence that is expected by most institutions of higher education. Taking an exam is perhaps one of the most independent activities a student experiences in a course, especially active learning courses, so this particular cultural difference may be exaggerated when it counts the most, during assessment.

This mismatch of norms is one reason that choices in course structure can impact different groups differently. For example, active learning has been shown to support learning for all demographic groups on average but has been disproportionally beneficial for historically marginalized populations \cite{paul_examining_2025, decaro_active_2025}. Similarly, adding retake exam options \cite{webb_attributing_2023, webb_highstakes_2025} and reducing the weight of the exam component \cite{Simmons2020} serves to mitigate the harmful effects of bias against women in introductory physics. And teaching concepts first benefits historically marginalized populations \cite{webb_attributing_2023, Webb2017}.

If researchers were to show that the correlation between uncompleted (blank) test items and student understanding was different for different demographic groups, then the practice of awarding a zero for an uncompleted response would on average hurt some demographic groups more than others and would then therefore be a cause of the equity gap. In this paper, we do just this.


\section{\label{sec:PriorPaper}Prior Work on Blank-Leaving Behavior}

In a previous conference paper \cite{Paul2018}, we analyzed the number of uncompleted test items by different demographic groups. We found that women left just over 3\% of problems uncompleted and that men left just under 3\% of problems uncompleted and that the difference between them was statistically significant. We also found that students belonging to groups that have been historically marginalized and first college generation students left more than 3.5\% of problems uncompleted. These differences might seem small, but when the percentage grade scale is used, the zeros earned by the blank responses can have a devastating impact on course grades \cite{Webb2020, Paul2022}. For example, Guskey \cite{Guskey2013} reminds us ``To recover from a single zero in a percentage grade system, a student must achieve a perfect score on a minimum of nine other [equal] assignments."

In our previous work \cite{Paul2018}, we used two separate analyses to determine if the demographic gaps in uncompleted test items were indicative of a differential lack of understanding of physics concepts. First, we plotted the average grade for completed answers by the fraction of answers left uncompleted for each student, and noted while there was a correlation between more uncompleted items and lower scores on completed answers, that correlation was weak ($R^2$ = 0.13) as the plot was extremely scattered. Second, we showed that the demographic differences in the number of uncompleted items existed even for high performing students.

The authors of this paper present data from the previous conference paper to audiences who are often surprised by the results. We have found from interactions with some past audience members that they make changes to their assessment strategies as a result of this research. For this reason, the authors have decided to revisit this work and prepare a more robust analysis for a wider audience.

\section{\label{sec:Framing}Framing}
As we have done in recent previous work \cite{webb_attributing_2023, paul_examining_2025}, we employ two models that guide us in making equitable changes as instructors, and examining the success of those changes as researchers. First, we use the Equity of Parity model \cite{Rodriguez2012} which defines an equitable course as one where students of each demographic group achieve on average the same overall outcome. In other words, course grades are not predicted by demographic characteristics. One difference in how we are using the Equity of Parity Model to what is described in Rodriguez et al. \cite{Rodriguez2012}, is that instead of gaps being the implied result of past inequities, we propose that measurement choices can also be the cause of gaps, and that it's possible to close the gaps entirely by addressing bias in measurements.

Similarly, we employ a course deficit model \cite{Cotner2017} which (when compared with a student deficit model) emphasizes that equity gaps are a characteristic of the course (which can be changed), and not a characteristic of the separate demographic groups resulting from prior experiences. 
Importantly, we are not saying that there are no differences in the experiences of different demographic groups. We are instead proposing that certain course structures privilege assets that are on average, more prevalent to  some demographic groups over the assets of others and, further, that these course structures may be changed to remove the biases and allow marginalized groups' abilities and/or achievements to be properly recognized and recorded.

\section{\label{sec:Questions}Research questions}

Giving a grade of zero for an uncompleted response can have a very negative impact on a student’s overall exam score, especially if their instructor is using a traditional form of the percent scale \cite{Reeves2004, Guskey2013, Webb2020} where earning somewhere around 50 or 60\% of the points is considered failing. In this paper we do not draw any conclusions on why students leave questions uncompleted. Instead, we evaluate the extent to which exams items for each student demographic group go uncollected. For this exploratory study, we consider a few broad groupings of students by gender, race and ethnicity, and first generation college status. Specifically we ask:

\begin{enumerate}
    
    \item How does the rate of uncollected item-level assessment data (blanks) vary by student population?
    \item Does the rate of uncollected item-level assessment data correlate with other measures of understanding for students?
     \item How does the distribution of uncompleted test items vary by class section?
\end{enumerate}

The first question examines whether or not uncollected test items present an equity concern. The second question examines the extent to which an instructor might be able to claim that uncollected test items represent a lack of understanding of the material as opposed to a failure of a test to collect meaningful assessment data. And our last question explores the extent to which uncollected test items are associated with the course section rather than with the behaviors of individual students.

We examine exam grades in two introductory college physics courses in the Collaborative Learning through Active Sense-making in Physics (CLASP) curriculum \cite{Potter2014Sixteen} over more than fifteen years and show that different populations of students engage in these behaviors at different rates, and provide evidence that these differences are not adequately explained by any lack of student understanding. We also show that the rate of uncollected test items is course dependent. Finally, we provide a couple different suggestions for instructors concerned with accurately assessing student understanding of physics, and therefore minimizing the chance of missing an opportunity to collect assessment data.

\section{\label{sec:Data}Database Details}

\subsection{\label{sec:GradeData}Exam-item grades}

In a prior study of differences between grade scales \cite{Paul2018}, we used 96 original class databases from the years 2003-2012 to compile a set of 794,088 grades given on individual parts of 606 quizzes and 76 final exams, to 15,207 students taking either of the first two courses in the CLASP series \cite{Potter2014Sixteen}. To these grade data we matched student’s self-reported demographic data that we received from university administrative sources. For the current paper we built a new dataset adding to this first set of classes all of the other anonymized class databases from this course series that we had access to and that followed one of the two well-defined grade scales used in these classes.  For each of the two grading scales (4-point scale and percent scale) we know that a grade of zero on an open response question tells us that the item was uncollected (blank). All of the course databases that we had access to were given between 1997 and 2014.  All together we have a grade database consisting of 1,290,528 grades given on individual parts of 951 in-term exams (mostly 30-40 minute long quizzes) and 134 final exams in 155 classes (60 classes from part A of the series, CourseI, 48 from part B, CourseII and 47 from CourseIII). Included are 20,281 unique students: 5,422 students appear three times (once in each of CourseI, CourseII, and CourseIII), 6,220 students appear twice (once each in two parts of the series), 8,179 students appear just once in the series, and 460 students appear more than once in at least one part of the series. 

These classes utilize the “Grading by Response Category” \cite{Paul2014} method, where student solutions are placed in categories that are later assigned a numeric grade. The 155 class-level databases include the individual grades that were given for each problem, the total exam scores for each student, the calculations that led to overall exam scores, and (in many cases) the calculations that led to a final grade.  The average number of exam item grades on an in-term exam is 3.6, the average number of exam item grades on a final exam is 15.3, and the average number of exam item grades for all exams over the term is 35.6.

We have examined copies of the original grading rubrics for 57 separate exams from the CourseI and CourseII classes over these years (this paper largely deals with the first two parts of the series) corresponding to about 200 exam problems, and found that 95\% of the problems for which a score of zero was possible assigned zero only to ``blank''or ``essentially blank'' solutions, meaning that there was no work shown evidencing that the student attempted the problem. In Appendix \ref{sec:datacleaning} we point out that we have removed about 2\% of the initial 3941 exam items from the CourseI and CourseII classes in the database because they are likely true/false types of questions where a zero indicates an incorrect answer rather than an answer left uncompleted. Our focus is only on open ended questions that have been uncollected (blank). We have also removed nearly 3\% of the initial exam items due to the grader deciding to use a grade method for that exam that, again, makes it unclear that a zero corresponds to an answer left blank (these graders usually calculated the exam grade by adding exam item grades rather than averaging them).  A more detailed discussion of the removal of exam items can be found in Appendix \ref{sec:datacleaning}. This appendix also includes our justification for leaving CourseIII out of our numerical analyses. For the results sections of the paper we will use only CourseI and CourseII with the major focus on CourseI and including CourseII results as somewhat independent measures of the same student characteristics, since we can see the same student in different courses.
CourseI class sections are very similar to each other, and CourseII class sections are also very similar to each other. Each course uses its own set of activities for 5 hours and 20 minutes per week, leaving only 1 hour and 20 minutes per week for standard lecture and assessment time. For more about the CLASP series courses see Potter et al. 2014 \cite{Potter2014Sixteen}.

\subsection{\label{sec:Demographics}Student demographics}

This university started collecting self-identified gender information in the Fall term of 2015 so our dataset is almost completely binary gender only. It includes 7,761 students identified as male, 12,433 identified as female, 2 identified as non-binary, and 85 whose gender is not listed. The university administration considers first-generation students to be those for whom neither parent had earned a bachelor’s degree or higher and we will refer to this group as 1stGen. We only have first-generation information for 62\% of the students in our dataset and, of those, there are 4,385 1stGen students in our dataset. Our dataset includes information on race/ethnicity as recorded by the university for 94\% of the students. To identify racial and ethnic groups that have been historically marginalized in STEM, we use the categorization from our previous work \cite{Potter2014Sixteen, Paul2017}. Students who self-reported belonging to any of the groups African American, Native American, Latina/o American, Mexican American, Chicana/o, or Pacific Island American are included in this group and hereafter referred to as HM for historically marginalized. There are 2,643 students in our dataset from these historically marginalized racial/ethnic groups.

\begin{figure} [htb]
\includegraphics[trim=2.2cm 1.0cm 2.2cm 1.0cm, clip=true,width=\linewidth]{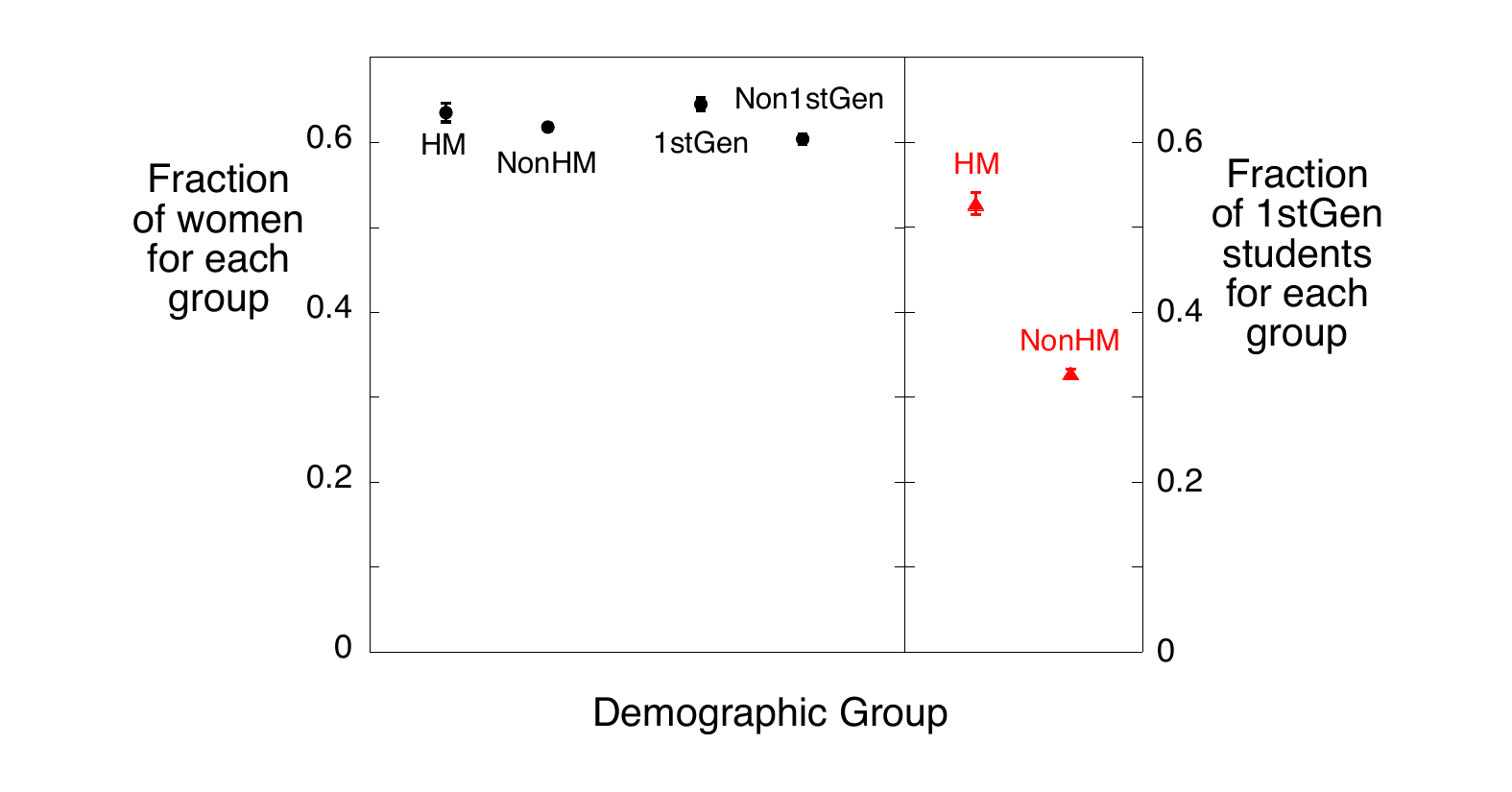}
\caption{On the left, the fraction of students who are women is shown, comparing historically marginalized (HM) and NonHM ethnic groups, and comparing first generation college students (1stGen) and Non1stGen. On the right, the fraction of students who are 1stGen is shown, comparing HM and NonHM ethnic groups.}
\label{fig:InterGroupDemog}
\end{figure}

We should note that gender is a distinctly different kind of demographic variable than HM status or 1stGen status. Those latter groupings are determined by variables that describe parents/guardians/extended-family of a student but gender is specific to the student and tells us nothing about parents/guardians/extended family. For example, the HM and 1stGen variables are slightly correlated with each other in our dataset while gender has almost no correlation with the other two groupings. We use our data to show this in Fig. \ref{fig:InterGroupDemog} where the left half of the figure shows that the fraction of women in our data is approximately constant over the other four demographic groupings. Conversely, the right half of the figure shows that the fraction of 1stGen students in the data does depend on which of the other two racial/ethnic demographic groups we are considering. We will also find differences in the probabilities of leaving blanks that may be related to the difference in the type of demographic grouping being considered.

\section{\label{sec:Results}Results}

\subsection{\label{sec:OverallDemographics}Demographics of uncompleted test items}

\begin{figure} [htb]
\includegraphics[trim=2.0cm 2.9cm 3.1cm 3.6cm, clip=true,width=\linewidth]{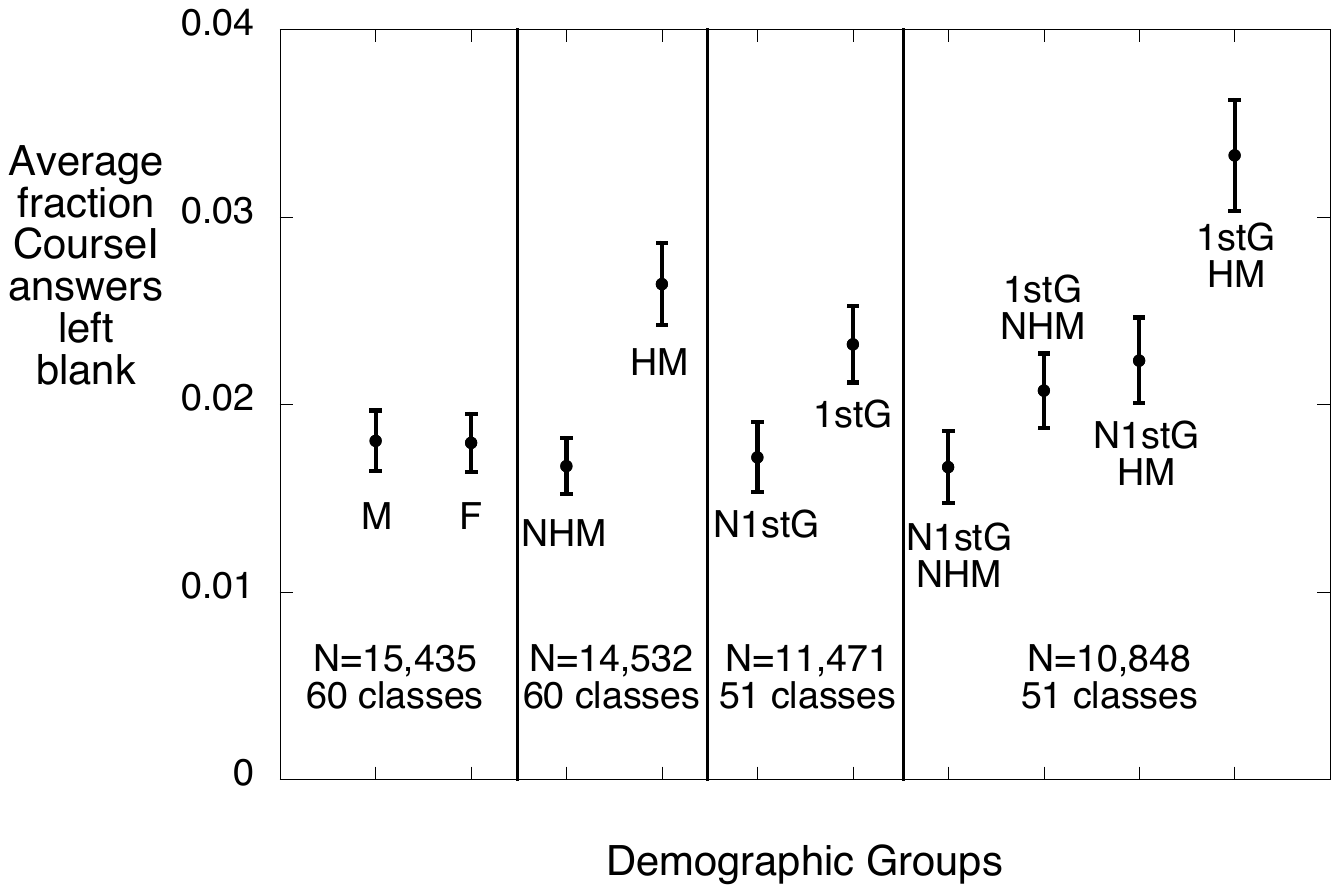}
\caption{The average fraction of uncollected item level responses in CourseI classes is plotted for demographic groups differentiated by gender, HM status, as well as 1st generation status. We also include intersections of HM with 1st generation. For each division of the students into demographic groups we give the total number of students for whom we have the appropriate identifying data and the number of classes over which we must average.}
\label{fig:FracBlanksInAAllDemographics}
\end{figure}

To answer research question \#1, we want to determine the rate of uncollected test items (blanks) from different demographic groups, but we first need to identify what kind of analysis is appropriate given differences between class sections. The fraction of blank answers on exams varies from class to class and covers the range from 0\% of answers left blank in an entire class's exams to more than 6\% of answers left blank in an entire class's exams. The fact that there is large variation of these numbers at both the student level and the class level leads us to use hierarchical linear modeling (HLM) to determine the appropriate averages and error estimates for each demographic group we consider.  Specifically, we use HLM to fit the overall average fraction of blanks in CourseI classes for each of ten demographically-defined groups.  The lowest level of the hierarchy is the student level and the second level consists of the classes (where groups of students took the same exams).  So we use HLM to fit the following equation: $FracBlanks = b_0$ and the resulting coefficient $b_0$ will be the appropriate average, first over students within a class and then over classes, of the fraction of blank answers ($FracBlanks$).  Two demographic groupings are defined by gender (M or F). Two are defined by historically marginalized race/ethnicity status, HM or NonHM (NHM). Two are defined by first-generation status (1stG or N1stG). Finally, four are intersections of HM status and 1stG status (N1stG and NHM, or 1stG and NHM, or N1stG and HM, or 1stG and HM). Note that we have first-generation status information for only 51 classes. Figure \ref{fig:FracBlanksInAAllDemographics} shows the average fraction of uncollected test items in each demographic group with standard errors that, because we are using a multilevel modeling, include both the variations between students and the variations between classes.

A more careful analysis than our previous work \cite{Paul2018} shows a more conservative estimate of the overall fraction of uncollected item level responses (blanks) from each demographic group. Nevertheless, the differences between students with HM status and their peers and between 1stGen students and their peers remains. On the other hand, the small gender difference seen in our previous paper \cite{Paul2018} has disappeared because the gender differences in the apparent True/False questions that we removed were the source of our earlier measured gender difference. In addition, compared to our prior work, the standard errors have increased due to our accounting for class-to-class variations. In Appendix \ref{sec:AllGenData}, we extend the gender result and show that, not only is there no overall gender difference, there is no measurable gender difference in uncollected test item responses over the set of the twenty-three individual (and overlapping) demographic groupings that we can define with the data we have on hand.  

Next, we analyze the consequence of the intersection of HM and 1stGen status and, in Fig. \ref{fig:FracBlanksInAAllDemographics}, show that the effects of being in more than one marginalized demographic group are compounding. For example, exam items are twice as likely to be uncollected for 1stGen students belonging to HM groups compared with non-1stGen students belonging to non-HM groups. Reframed from a course deficit perspective: the exams are failing to capture data necessary to evaluate students from historically marginalized groups who are also 1st-Gen students. Appendix \ref{sec:CourseB} shows a similar result holds in CourseII.

For completeness, we note that these demographic disparities in the fraction of missing data include disparities both in the numbers of students for whom data is missing as well as the amount of data that is missing.  To show this we use the four groups formed by intersections of HM and 1stG, and find the fraction of each group that left either zero, one or two, and more than two blanks during the term.  These fractions are shown in Table \ref{tab:DemogrFracLeavingBlanks}.  We see that, when compared to the non-marginalized students, an extra 6\% of the most marginalized students leave one or two answers blank during a term.  But, for more than 2 answers blank, that extra number amounts to about 8\%.
\begin{table}[htbp]
\caption{The percentage of students from each single demographic group who left either zero, one or two, or more than two answers blank over the course of a full term.  Standard errors are in parentheses.  Marginalized students have larger numbers of data missing at higher rates than their peers.}
\label{tab:DemogrFracLeavingBlanks}
\begin{ruledtabular}
\begin{tabular}{c c c c}
\textbf{Group} & \textbf{0 Blank} & \textbf{1or2 Blank} &\textbf{$>$2 Blank}\\ 
\hline
1stG/HM & 54.3 (1.4)\% & 31.6 (1.3)\% & 14.1 (1)\% \\
N1stG/HM & 62.3 (1.4)\% & 29.0 (1.3)\% & 9.6 (0.9)\% \\
1stG/NonHM & 62.6 (0.7)\% & 29.4 (0.6)\% & 8.0 (0.4)\%  \\
Non1stG/NonHM & 68.5 (0.5)\% & 25.8 (0.4)\% & 5.7 (0.2)\% \\
\end{tabular}
\end{ruledtabular}
\end{table}

\subsection{\label{sec:HighPerformingStudents}Are the demographic patterns we see driven by low performing students?}

\begin{figure} [htb]
\includegraphics[trim=2.2cm 2.9cm 2.7cm 3.9cm, clip=true,width=\linewidth]{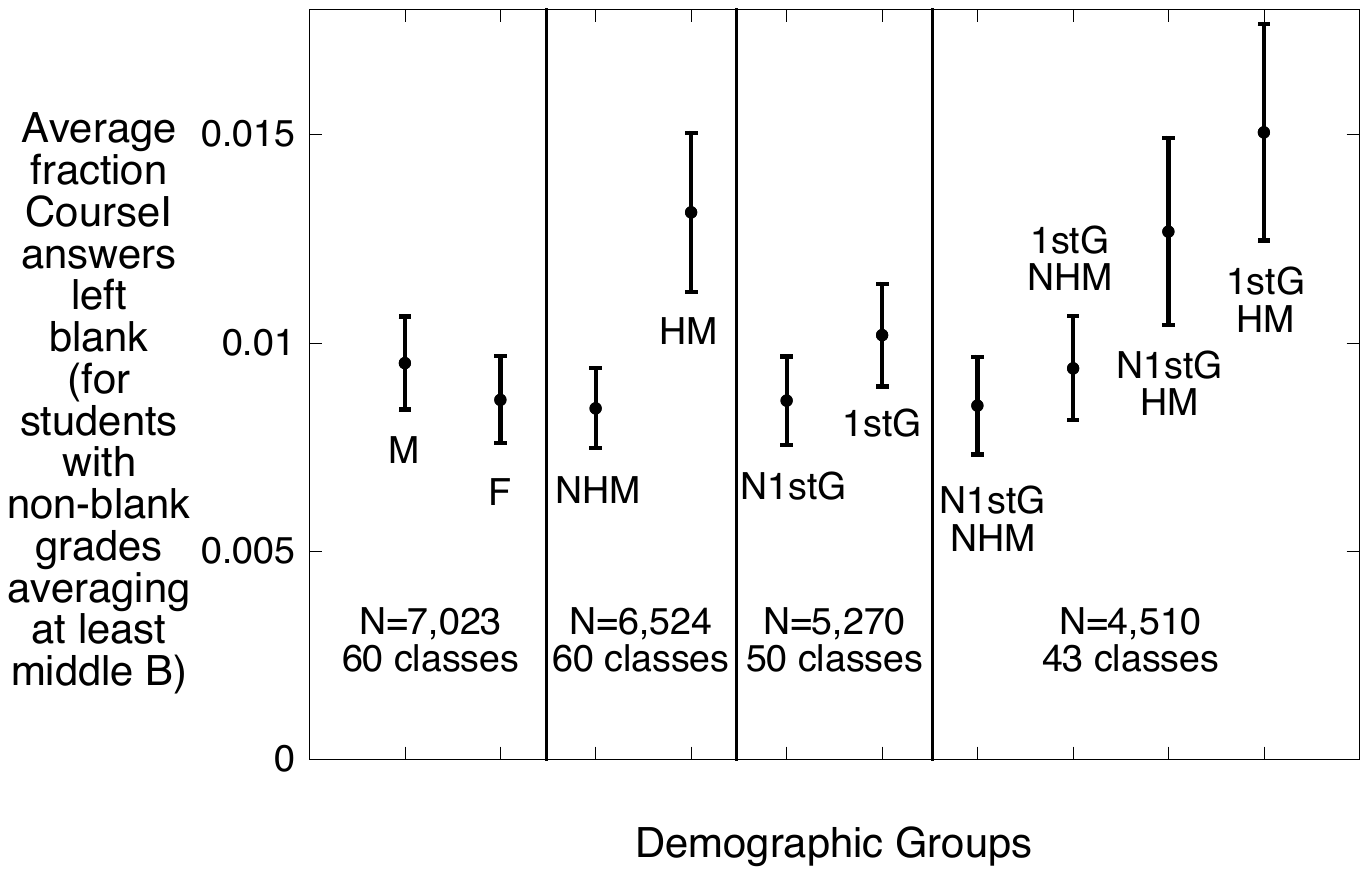}
\caption{The average fraction of uncollected item level responses (blanks) in CourseI classes for students, whose non-blank answers average to a middle B or better, is plotted for demographic groups differentiated by gender, HM status, as well as 1st generation status. We also include intersections of HM with 1st generation.  For each division of the students into demographic groups we give the total number of students for whom we have the appropriate identifying data and the number of classes over which we must average.}
\label{fig:FracBlanksInABGradeDemographics}
\end{figure}

As a way of showing that demographic differences are not limited to students with low grades in the course, we limit the set of students to those whose average of completed test items (non-blank grades) was equal to a middle B or higher.  This set of students is a little over 40\% of the total dataset.  The grade cutoff not only limits the number of students per class but also begins to limit the number of classes in which there are students within each comparison group who meet the cutoff.  The results are shown in Fig \ref{fig:FracBlanksInABGradeDemographics}.  One notices that the average number of uncollected test items from students who are in both group HM and 1stG is about twice the average number uncollected from students who are in neither of those two groups. This holds both for this subset of students doing well in the course, Fig. \ref{fig:FracBlanksInABGradeDemographics}, and for the entire set of students in the dataset, Fig. \ref{fig:FracBlanksInAAllDemographics}, suggesting that differences in uncollected test item rates don't necessarily represent differences in ``demonstrated understanding''.  This conclusion then supports the idea that the uncollected items themselves may not always represent a lack of understanding. At this point it seems relevant to note that, as we have shown in previous work \cite{webb_highstakes_2025}, the exam scores of these students likely include poorly understood demographic biases.  These biases lead to different grades for different demographic groups even when each group's average understanding/skill is the same.  So using a grade cutoff (which include these biases) as we do in Fig. \ref{fig:FracBlanksInABGradeDemographics} results in group comparisons that are somewhat ill-defined.

For completeness, in Appendix \ref{sec:AllEthnicites} we compute the average fraction of uncollected exam item responses for each of the racial/ethnic groups that have a significant number of students in our database and find that the racial/ethnic groups who, on average, are more likely to have uncollected test items are the groups making up HM and, in addition, Korean-American students.

\subsection{\label{sec.FracAsCorrelations}How do uncollected test item responses correlate with high grades and low grades?}

The fraction of A-grades that a student receives on exam answers is certainly a measure of how well they are doing in the class and simple addition (i.e. total fraction of grades = 1) tells us that the fraction of A-grades ($FracAs$) a student earns for their answers on exams is likely to be inversely correlated with the fraction of blank answers ($FracBlanks$) and also with the fraction of D and non-zero F grades ($FracDsOrFs$), and, of course, positively correlated with the grade in the course ($CourseGrade$), so let’s examine these correlations.

Table \ref{tab:Correlations} shows the Pearson correlation coefficients for $FracAs$ with $CourseGrade$, $FracDsOrFs$, and blank answers $FracBlanks$ from the CourseI classes.  All three signs are as expected but, though $FracAs$ has the expected relatively large correlations with $CourseGrade$, and $FracDsOrFs$, $FracAs$ has a much weaker correlation with $FracBlanks$. The CourseII classes, Appendix \ref{sec:CourseB}, have a similar set of correlations.  The fact that the correlation between $FracAs$ and $FracBlanks$ is much weaker than that between $FracAs$ and $FracDsOrFs$ suggests that a student's decision to leave a blank is not simply showing that they don't know how to work the problem.  $FracDsOrFs$ represents a graders evaluation of an answer so it is not surprising that it is highly negatively correlated with $FracAs$ which also represents that grader’s evaluations.  The uncollected data, represented by $FracBlanks$, is not nearly as negatively correlated with $FracAs$ as the D and F grades suggesting that the uncollected data differs in some way(s) from answers that have been evaluated by a grader.

\begin{table}[htbp]
\caption{The correlation coefficients between the fraction of exam-item A-grades given to a student from CourseI and three other student-level grade variables. The fraction of A-grades is, as expected, strongly correlated with the course grade and the fraction of low grades (Ds or Fs) but has a much weaker correlation with the fraction of answers left blank.}
\label{tab:Correlations}
\begin{ruledtabular}
\begin{tabular}{c c c}
\textbf{Grade} & \textbf{Correlation between} &\textbf{p-value}\\ 
\textbf{variable} & \textbf{variable and FracAs} &\\ 
\hline
$CourseGrade$ & 0.827 & $<10^{-4}$ \\
$FracDsOrFs$ & -0.737 & $<10^{-4}$ \\
$FracBlanks$ & -0.308 & $<10^{-4}$ \\
\end{tabular}
\end{ruledtabular}
\end{table}

\subsection{\label{sec.EffectsOfClass}Are uncollected test items more associated with the student or the course section?}

Including all classes in our database, the fraction of uncollected exam answers is not very different for CourseI of the course series (1.79\% $\pm$ 0.15\%) and CourseII (1.71\% $\pm$ 0.20\%). These two numbers actually involve a large fraction of the students, 35.1\% in CourseI and 34.2\% in CourseII left at least one response blank. The similarity between these numbers might lead to the idea that perhaps 1/3 of the students are willing to leave an answer blank and other 2/3 are not and that blank-leaving is largely determined by the student. This simple idea gets clouded somewhat when we see that, for the subset of 8,793 students for whom we have both CourseI scores (where 37\% left at least one blank) and CourseII scores (where 36\% left at least one blank), the fraction of students who left at least one blank in \textbf{one} of the \textbf{two} courses is actually 57.5\%, a majority of the students, indicating instead that the majority of students are willing to leave a blank in some circumstances. 

\begin{figure} [htb]
\includegraphics[trim=4.0cm 2.4cm 5.1cm 3.6cm, clip=true,width=\linewidth]{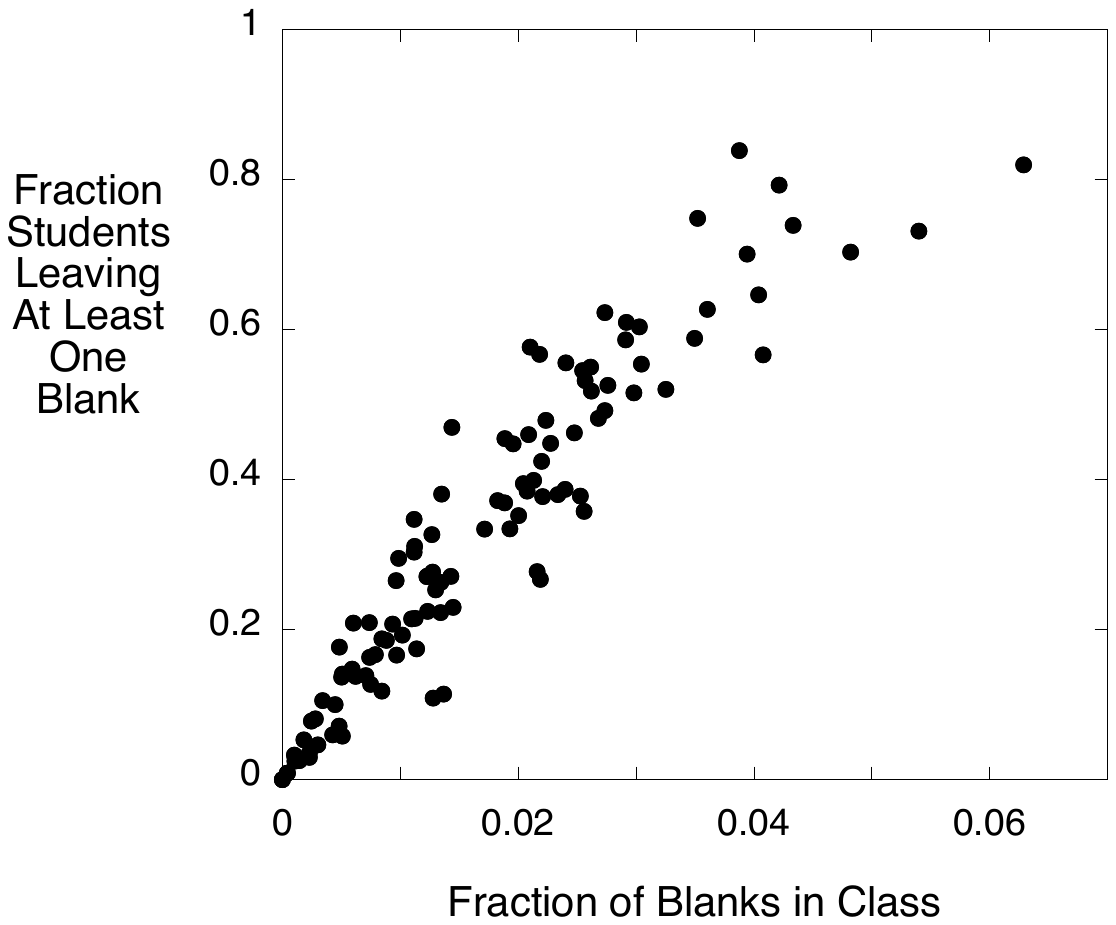}
\caption{The fraction of students in a class who left at least one blank is shown for all 108 CourseI and CourseII classes in our database. These fractions are plotted against the student-average of the number of blanks left on that class's exams.}
\label{fig:FracStudLeaveBlank}
\end{figure}

We can get a more complete picture of this phenomenon of blank responses by looking at the variations of these numbers over all 108 classes. As a way of displaying this class-dependent variety of blank, Fig. \ref{fig:FracStudLeaveBlank} shows, for each of the 108 classes (from CourseI and CourseII of the course series), the fraction of students who left at least one blank sometime during the term as a function of the student-average total number of blanks left for that class. In 11 of these classes fewer than 10\% of the students left a blank the entire term and, on the other end, in 11 of these classes more than 74\% of the students left at least one blank during the term. Since these classes are essentially identical except for the teachers and the exams, it seems likely that the specific set of exams plays a large part in determining if a student will leave an answer blank. These data suggest that the course itself plays a large part in determining how much data goes uncollected.

We saw that almost 58\% of students left a blank answer in at least one of their first two physics courses. One way to quantify the student-effect on blank leaving is to compare blank-leaving in the students CourseII work with blank-leaving in their CourseI work to see if the same set of students is leaving blanks in each of their classes. For one look at this situation we can just calculate the correlation between students who left at least one blank in CourseI and those who left at least one blank in CourseII. We find that there is very little correlation between these numbers, Pearson's correlation = 0.11 (P $< 10^{-3}$).  A comparison between the fraction of blanks in CourseI and the fraction in CourseII yields a similarly weak correlation of 0.19 (P $< 10^{-3}$). The fact that blank-leaving behavior in a physics course is not well correlated with blank leaving in the next physics course does not give us confidence in the idea that blank-leaving is largely controlled by the student.  Instead we reframe these phenomena as courses failing to collect evaluation data from the students.

\subsection{\label{sec:EffectsBlanks}Effects of uncollected test item responses on course grades}

About 65\% of the students in CourseI left no answers blank. The 35\% who left at least one blank left an average of 1.87 blanks on an average of 37 exam items during the term. We can use our database to estimate the average effect on a student's course grade when they were one of the 35\% who did leave at least one blank. We'll use the variable $LftBlnks$ = 1 for a student who left at least one answer blank and = 0 for a student who left no blanks. We'll control for physics knowledge/skills so that we are comparing students who left blanks with equivalently skilled students who chose not to leave any answers blank. We have already seen that the fraction of exam-item grades in the A-range ($FracAs$) is well correlated with the course grade. In addition, in a previous paper \cite{Webb2020} we showed that the course grade has a small quadratic dependence on $FracAs$ so we will use both the linear and quadratic terms to control for demonstrated achievement in the course (just using the linear term leads to the same results and conclusions). We will also include the grade scale, $GrdScl$ = 0 for 4-point grading and = 1 for percent scale grading, because we know from our prior work that the grade scale has a measurable effect on the course grade due to the averaging effects of the extra low F-grades in a percent scale system.\cite{Paul2022, Webb2020}  Finally, for the same reason that we include the grade scale, we include an interaction term between $GrdScl$ and $LftBlnks$. We again use Hierarchical Modeling and fit the following model
\begin{equation}
\begin{split}
CourseGrade & = b_0 + b_{FracAs}FracAs \\&+ b_{FracAs^2}FracAs^2\\ &+ b_{LftBlnks}LftBlnks \\ &+ b_{GrdScl}GrdScl \\&+ b_{GrdScl*LftBlnks} GrdScl*LftBlnks
\end{split}
\label{eqn:GrakeLeftBlanks}
\end{equation}
where the coefficient $b_{LftBlnks}$ will measure the average effect on course grade of being in the group who left one or more answers blank in a 4-point-scale graded class and $b_{LftBlnks}$ + $b_{GrdScl*LftBlnks}$ will measure the same grade differential but for a student leaving blanks in a percent-scale graded class.

\begin{table}[htbp]
\caption{The coefficients from an HLM fit to Eq. \ref{eqn:GrakeLeftBlanks} are shown along with their standard errors, z-statistics, and p-values.  Included are the 107 classes from the CourseI and CourseII parts of the series, for which we have course grades, with a total of 27,008 students.  The coefficient $b_{LeftBlanks}$ measures the average effect on a student's grade resulting from that student being in the group who left at least one blank.  That group left an average of 5.25\% answers blank.}
\label{tab:GradeEffectBlanks}
\begin{ruledtabular}
\begin{tabular}{c c c c c}
\textbf{Coeff.} & \textbf{Value} &\textbf{Error} & \textbf{z-statistic}
& \textbf{p-value}\\ 
\hline
$b_{FracAs}$ & 4.584 & 0.048 & 95.47 & $<10^{-3}$ \\
$b_{FracAs^2}$ & -1.336 & 0.052 & -25.49 & $<10^{-3}$ \\$b_{LftBlnks}$ & -0.0918 & 0.0063 & -14.52 & $<10^{-3}$ \\
$b_{GrdScl}$ & -0.086 & 0.046 & -1.85 & 0.064 \\
$b_{GrdScl*LftBlnks}$ & -0.218 & 0.011 & -20.66 & $<10^{-3}$ \\$b_{0}$ & 1.392 & 0.030 & 47.0 & $<10^{-3}$ \\
\end{tabular}
\end{ruledtabular}
\end{table}

Table \ref{tab:GradeEffectBlanks} shows that leaving blanks under a 4-point grading regime led, on average, to a lower grade by little less than a tenth of a grade point. On the other hand, under a percent-scale grading regime leaving answers blank resulted, on average, in over three times larger grade drop, about 0.3 grade points (for example from a B to a B-). The arithmetic of averaging grades, when blanks are given zeros, makes it obvious that the percent-scale is a harsher grade scale and this is born out in practice We find similar results if we measure the grade difference for students leaving only one answer blank during the term (instead of also including students who may have left two or more answers blank).  We should note that in essentially all of the classes in our dataset the course grade is calculated after dropping each students' lowest exam.  

\section{\label{sec:Discussion}Discussion}
In our prior work \cite{Paul2018} we claimed that because students from different demographic groups left different numbers of physics problems blank, and because blank-leaving behavior wasn't entirely explained by metrics indicating lack of understanding, that blank-leaving behavior is effectively missing-data and should be treated as such in the grading process. 
In this paper, using a course deficit model we reframed this same phenomena: Exam items are uncollected from different demographic groups at different rates and this means that measurements of students skills are, on average, biased against some groups.


We add to our prior work in several important ways. At the individual student level, we investigated the correlation between uncollected test item responses and metrics for understanding in a new way, and came to the same conclusion that the number of uncollected items per student is not adequately explained by lack of understanding. 

A more careful analysis of gender differences revealed that our previous finding that test item responses are more likely to be uncollected from women as false. Our data show that on average, men and women leave the same number of blanks on exams. Interestingly, when we account for statistical anomalies, this lack of gender gap seems to hold true across all races/ethnicities we are able to measure it for.

We continue to find that two groups, first-generation students and students who are members of historically marginalized ethnic groups, had higher rates of uncollected test item responses. In addition, we find that belonging to more than one of these marginalized groups has a compounding effect on the likelihood of uncollected test item response rate. This finding is important for future quantitative equity related studies in that it illustrates that exploring the impact of course structures on students may be inadequate if studied only from a single dimension of race, gender, or other status. 

In Appendix \ref{sec:AllEthnicites} we show that some racial/ethnic groups are more likely to have uncollected test items than others. While the data in this paper does not point to particular reasons why test item responses are uncollected at different rates across demographics, one interesting possible interpretation is related to the finding that gender is not a factor that determines the likelihood of whether test item responses are collected while familial variables (race/ethnicity and first generation status) do. Students' values associated with their family background may influence the likelihood of a particular student leaving a blank or not. 
The following quote from a student on the practice of blank-leaving and partial credit can help illustrate this phenomenon:

“\textit{Physics is not bullshit. You shouldn’t be writing bullshit answers, but you have to. I don’t know, it feels a little dirty but at the same time I want a good grade on an exam. . . . Physics is not based on bullshit but here I am.}”, page 70 of M.K. Chessey's dissertation \cite{Chessey2018}.

In the above quote the student expresses that they would prefer not to write down things they are unsure of when presented with a challenging problem under time pressure, but this is exactly what is expected of them in this particular environment. In this example, the student is very reluctant to sacrifice their value of submitting only high-quality work in order to get ``a good grade.'' Most instructors would agree that this is a good value for students to hold. However, it is also a value that is penalized during most timed in-person exams. Similarly, this practice of ``bullshitting'' might make this student feel less like a scientist because they didn't feel they deserved the points they earned for partial credit. Depending on their values, different students might be more or less likely to leave a blank in a given scenario. Perhaps these values are associated with familial background and this is why we don't see a difference in gender.

We want to stress here that blank-leaving behavior is not necessarily a good or bad thing, as our quote above illustrates that there is a blank-leaving vs. `bullshitting' tension that might be best considered from a values perspective that is beyond the quantitative scope of this paper. Speirs et al. have recently done important work \cite{speirs_thematic_2026} cataloging student feelings around grades and further research is needed here to understand the extent to which exams and grades shape the learning experience and outcomes beyond their numerical impact.

Perhaps most importantly from an intervention standpoint, we find that some course sections in our sample (despite having similar assessment types - in this case timed in-person exams/quizzes) have drastically fewer uncollected test item responses than others. We also show that if a student leaves a blank in one course this does not correlate with them leaving a blank in another course. This indicates that uncollected test items (or blanks) are a variable that is attributed to course structure thus suggesting (as we have done in prior work \cite{webb_attributing_2023, paul_examining_2025, webb_highstakes_2025}) that a Course Deficit Model \cite{Cotner2017} rather than a student deficit model, is an appropriate way to consider equity gaps in higher education. Furthermore, because we see examples where course sections have similar assessment formats but drastically different amounts of uncollected test item responses, we infer that it is possible to increase the rate of collected test items from student, and thus have more data to effectively evaluate students. Future work will examine the attributes of quizzes and exams (i.e. exam length, exam format, exam complexity) which are correlated with the rate of blanks.

As instructors we want students to demonstrate the full extent of their understanding on exams so that each student can receive a grade that best represents their knowledge and skill in the subject, but we can also see the issue from the point of view of a student who does not want to pretend to possess knowledge or skill that they feel they do not have. This is only one possible reason students may leave blanks, but it illustrates the need to examine grading student behaviors that are not necessarily representative of their knowledge, especially since exam items are more uncommonly uncollected from some groups of students than from others. 

Our finding that uncollected test item responses are course dependent is a positive development in the sense that it means that instructors who wish to collect assessment data more fairly have some options. While alternative grading strategies have been slow to be adopted in higher education, and research on these methods is lagging \cite{hackerson_alternative_2024} there are nevertheless several grading/assessment practices that could prove useful. A possible solution is to adopt mastery type grading styles like standards-based grading \cite{Beatty2013}, specifications grading \cite{leslie_specifications_2020}, or other type of mastery-based grading \cite{armacost_using_2003}. This is because these methods do not average student attempts on the same content. If a student leaves a blank, they have an opportunity to replace that score with another one. For the same reason, offering a re-take option on exams could also prove useful \cite{webb_highstakes_2025} Another option is to employ the practice of `minimum grading' \cite{Carey2012} which involves increasing low grades to 50\%. Using minimum grading means that uncollected exam items have a smaller impact on overall grades. Another option that allows graders to assign zeros to uncollected test item responses but weighs their impact more similarly to other letter grades, is to use a GPA-like 4-point scale \cite{Webb2020, Paul2022}. Use of the 4.0 (or 4.5) scale is shown in section \ref{sec:EffectsBlanks} to be an effective way to minimize the impact of uncollected test items. A short overview of some of these methods is provided by Townsley and Schmid \cite{Townsley2020} and may provide a nice entry point for instructors considering alternatives to traditional grading.

We urge the reader to frame the blanks issue not as as a problem of the students failing to complete an item, but instead as the problem being that the traditional exams are failing to capture data from students necessary to evaluate their skill. It is very tempting to try to address this problem from a perspective of teaching the students not to leave blanks. For example practices that promote growth mindset, or address stereotype threat, could possibly reduce the numbers of blanks made by students, and these are also excellent practices that should be used in classrooms regardless. However, this approach necessarily uses a student deficit model because it is trying to address a structural problem by changing the students and not the assessment. A course deficit model approach would change the assessment or grading in order to minimize or remove the bias.

\section{\label{sec:Limitations}Limitations}
One major limitation of this work is that we don't have an explanation for why test item responses are uncollected from different demographic groups at different rates, and why this effect can be compounding for students who belong to more than one demographic group. We also don't have access to all demographic groups that might make a differences. For example, students' socioeconomic background might also correlate with uncollected test item number, but we don't have data on that. Future qualitative work will investigate the choices that students make during exams, and the exams structures that influence these decisions, and this will hopefully help explain why there is variation between groups. 

Similarly, it is not yet known why some course sections have higher rates of uncollected exam item responses than others. Current hypotheses are that the exams in some classes are too long or too difficult, causing students to skip problems as a test-taking strategy. However, this alone would not explain the differences in demographic behaviors. In the future, we will examine the characteristics of exams with many blanks to provide further recommendations for instructors wishing to decrease blank-leaving outcomes. 

\section{\label{sec:Conclusions}Conclusions}
In this work, we find that exam item responses are uncollected from students belonging to different demographics (race/ethnicity and first generation status) at different rates.  For example, the rate of uncollected exam item responses from students who are not first generation and not from historically marginalized groups is about half as many  ($\sim1.5\%$) as students who are first generation and from a historically marginalized group ($\sim 3.0\%$). This same difference of roughly a factor of 2, exists even among students with a B average on completed test items, suggesting that blank leaving is not explained by a general difference in understanding among different populations of physics groups. Interestingly, there is no significant difference in uncollected test item responses by gender for any of the ethnic/racial demographics, possibly suggesting that differences in uncollected exam item data might be related to student values that are familial and/or cultural. 

In addition, we find that belonging to both historically marginalized and first generation college student groups has a compounding effect on the likelihood of uncollected test item response rate. This finding is important for future quantitative equity related studies in that it illustrates that exploring the impact of course structures on students may be inadequate if studied only from a single dimension of race, gender, or other status.

We also find that uncollected exam item responses are highly influenced by the course section. For example, the percent of students leaving a single problem blank in a single section of a course ranges from 0\% to 80\%. We also find that there is no statistical relationship between individual student blank-leaving behavior in CourseI (the first course in the series) and the blank-leaving behavior in CourseII (the second course in the series). Instructors who want to minimize inequities caused by uncollected exam item responses (blanks) should either consider approaches that minimize possibilities for blank leaving (allowing re-takes on exams, giving take-home exams, or creating shorter exams) or use grading practices that minimize the impacts of blanks (like standards-based grading, or the 4.0 scale). Some of these options are discussed in section \ref{sec:Discussion}.

\section{\label{sec:Acknowledgments}Acknowledgments}
The authors would like to thank the San Jose State University Physics Education Research group for feedback on this work as it was in progress. The National Science Foundation provided partial support of this project through the now terminated NSF HSI IUSE 1953760 grant. This work would not have been possible without the support of San José State University’s Research, Scholarship, and Creative Activity Assigned Time Program and the Research and Innovation Scholarly Entrepreneurship (RAISE) award, which allowed us to continue investigations on historically marginalized student populations after National Science Foundation funding was pulled from these efforts.

\appendix

\section{\label{sec:datacleaning}Data collection}

We have access to 279 anonymized grading databases from the total of 297 CLASP classes given between 1997 and 2014.  We’re limited to these years because in 2012 course instructors started keeping their own grade databases and the number that we had access to decreased to zero by 2015.  We narrow these down to classes where the instructor has used a well-defined grade scale so that we know that a grade of zero means that no answer was attempted and so that we know the letter-grade meaning of the non-blank grades.  Of these 279 we find 155 where the instructor used one of the two clearly recognizable grade scales, either percent scale grading or 4-point scale grading, and the in-term exam grading is, as far as we can tell, completely recorded.  The grade scale is identifiable because each exam item is given a grade between zero and the maximum (either 10 or 4.5 for percent scale and 4-point scale, respectively), each exam score is calculated using a weighted average of the exam item scores, and, when final exam scores are included, the calculation of course grades is also included in the database and matches the expected letter grades.  These 155 class databases include 60 databases from the first term in the series (CourseI), 48 databases from the second term in the series (CourseII), and 47 databases from the third term in the series (CourseIII).  We have removed 164 of the 27,032 sets of student grades because the student did not have grades on at least half of the graded questions (24 of these students were still given course grades).  This arbitrary cutoff doesn't change any of our numbers or conclusions.  We will discuss and compare the CourseI and CourseII databases in this paper because they are similar.  On the other hand, CourseIII databases are problematic in a way that we will discuss later in this appendix after we finish with CourseI and CourseII.  The analyses discussed in our paper are largely focused on CourseI databases with CourseII results referred to largely in support of CourseI results.

Even though we can positively identify the grade scales in these 155 classes there are still a few sets of exam item grades or even whole exams that don’t follow the grade scale of the rest of the database.  First, an entire exam might have exam items that are each given grades between zero and some maximum that is significantly less than the maximum of the grade scale.  The exam score on these exams was then calculated by summing the exam item scores rather than averaging them.  For these exams, we can’t count on a zero being a blank and we don’t have any way of assigning letter grades to the other scores so we removed them from our database.  For CourseI and CourseII together, we found that 100 out of 3941 sets of exam item grades had this situation and so we removed them.

Second, and more important, we found 76 of the 3941 sets of exam item grades had only two grades given to the students, maximum or zero.  We suspect these of being either True/False type questions or multiple choice type questions.  In either case the zero for these items likely means that the answer was incorrect rather than that the student left it blank.  We removed these exam item grades also because we want to only count answers left blank.  These particular exam item grades also had the largest fractions of zeroes (47 of them had more than 15\% zeros including 25 with more than 50\% zeros).  In our previous paper we had not removed these suspected True/False questions and their large fractions of zeros led to our estimated fractions of blanks being slightly higher than the data presented in this paper.  However, the differences between different demographic groups were mostly unaffected, so our main conclusions are largely the same as previously.

\begin{figure} [htb]
\includegraphics[trim=3.4cm 2.6cm 5.8cm 3.6cm, clip=true,width=\linewidth]{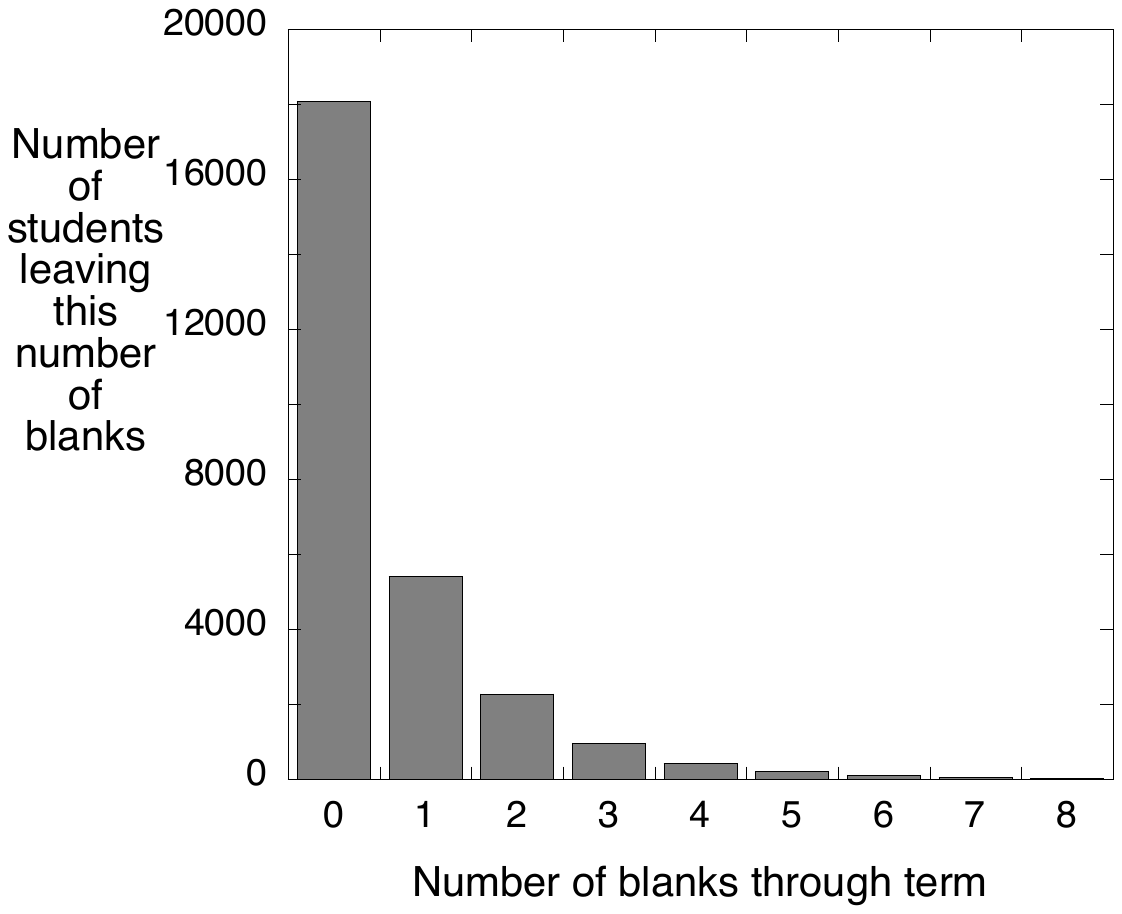}
\caption{Number of students vs number of blank answers they left during the term in courses I and II.  Although too small to show, there are also a few students leaving more answers, up to 15 answers, blank}
\label{fig:DistBlanks}
\end{figure}

In Sec. \ref{sec:GradeData} we noted that the average number of exam items in a term was 35.6 but that number varies from class to class, in our dataset, from a minimum of 11 exam items to a maximum of 72 exam items during the term.  This is our main reason for dealing with the fraction of answers left blank rather than the actual number left blank.  For completeness in the database discussion, Fig. \ref{fig:DistBlanks} shows the distribution of the actual number of blanks left during the term in courses I and II.

Finally, the CourseIII databases include exam items that seem likely to be giving zeros to wrong answers, rather than just to answers left blank, but which can’t all be easily identified and removed.  We suspect this for two reasons.  First, the average fraction of blank answers is somewhat larger for CourseIII than for CourseI and CourseII which were the essentially the same.  Second, and more telling, examination of some CourseIII exams (we have copies of some of them) shows us that there are some questions that were routinely asked in CourseIII that are essentially fill in the blanks type questions with several blanks to fill in.  In these questions filling in all blanks correctly led to a maximum score and filling some in incorrectly led to lower scores so that filling in none of the blanks correctly led to zeros even though the student attempted the answers (copies of CourseI and CourseII exams did not show evidence of any multiple-choice type questions).  We don’t have copies of all of the exams in these courses and there are no clear “fingerprints” in the data files themselves so we will not include the CourseIII classes in our analyses except to point out that the fraction of blank answers again increases with the number of marginalized identities and has no clear gender gap.

\section{\label{sec:AllGenData}Gender Comparisons within each other demographic group}

In Sec. \ref{sec:Demographics}, we noted that two of our demographic variables, HM status and 1stGen status, are variables associated with a student's family/guardian/extended family and we showed that the variable $Female$ had essentially the same average across those family-related variables.  In this appendix we'll provide the measured gender gaps in blank-leaving for each of the nine groups that we can define from the two family-related demographic variables.  To estimate the gender difference in blank leaving we define the variable $Female$ = 1 for women and = 0 for men and use HLM to calculate the coefficients in the following model:
\begin{equation}
FracBlanks = b_0 + b_{Female}Female
\label{eqn:GenderFracBlanks}
\end{equation}
where $b_{0}$ is the fraction left by men and $b_{Female}$ is the difference from that fraction for women.  We will fit this equation for each of the nine groups that we can define from the two family-related variables.  The nine groups include a group that leaves our entire dataset undivided, two more groupings will divide our dataset using the variable $HM$ = 1 if the student is from an historically marginalized racial/ethnic (HM) group and = 0 if they aren't (NonHM).  These two groups average over $1stGen$.  The rest of the nine groups are defined in a similar fashion.  The results of the fits are shown in Table \ref{tab:AllGroupGendGaps} and suggests that none of these possible demographic divisions of the group of students has a significant difference between the fraction blank answers left by women in the group and the fraction of blank answers left by men in that group.

\begin{table}[htbp]
\caption{The coefficients, $b_{Female}$, determined from HLM fits to Eq. \ref{eqn:GenderFracBlanks}, for each of nine groups with different demographic definitions, are the blanks gender gaps for those groups.  Each group is determined by the value(s) of either none, one, or both of the two family-related demographic variables (HM for students from historically marginalized racial/ethnic groups and 1stGen for students who are first generation college in their families).  The number of students in a group is N.}
\label{tab:AllGroupGendGaps}
\begin{ruledtabular}
\begin{tabular}{c c c c c c}
\textbf{HM} & \textbf{1stGen} & \textbf{GendGap} & \textbf{Error} & \textbf{P-value} & \textbf{N} \\
\hline
- & - & 0.00012 & 0.00053 & 0.826 & 15,414\\
0 & - & -0.00045 & 0.00055 & 0.410 & 12,496 \\
1 & - & 0.0012 & 0.0019 & 0.526 & 2,014 \\
- & 0 & -0.00029 & 0.00070 & 0.685 & 7,433 \\
- & 1 & -0.0014 & 0.0012 & 0.245 & 4,004 \\
0 & 0 & -0.00026 & 0.00074 & 0.729 & 6,284 \\
0 & 1 & -0.0024 & 0.0012 & 0.049 & 3,075 \\
1 & 0 & -0.0008 & 0.0030 & 0.792 & 693 \\
1 & 1 & 0.0013 & 0.0033 & 0.701 & 781 \\
\end{tabular}
\end{ruledtabular}
\end{table}

\begin{figure} [htb]
\includegraphics[trim=1.9cm 2.7cm 2.9cm 3.6cm, clip=true,width=\linewidth]{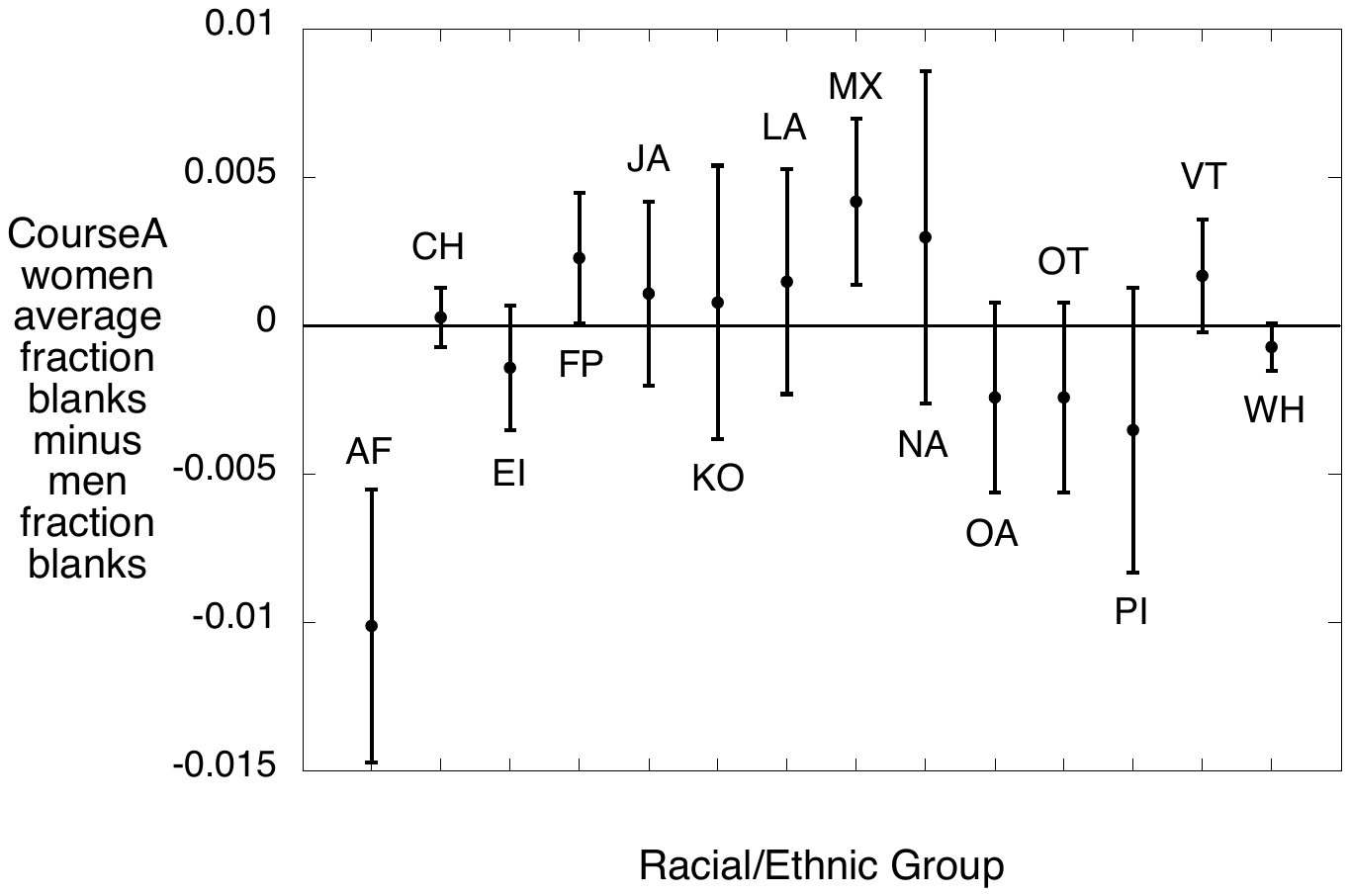}
\caption{The gender gap is, again, $b_{Female}$ obtained by fitting Eq. \ref{eqn:GenderFracBlanks} to CourseI data from each racial/ethnic group  shown in Table \ref{tab:tabA1}.  Assuming that there is \textbf{no gender difference for any racial/ethnic group} we note that for the 14 independent measurements shown in the figure there is higher than 30\% probability that one of the measurements will be found as far from zero as one sees for group AF and so we conclude that there is no reliable evidence of a gender effect.}
\label{fig:GndrGapByRaceCourseA}
\end{figure}

We can extend this analysis of gender gaps in CourseI to each of the individual racial/ethnic groups in our dataset.  These racial/ethnic groups are defined in Table \ref{tab:tabA1} and the gender gaps are shown in Fig. \ref{fig:GndrGapByRaceCourseA}.  As noted in the figure caption, assuming that there is no gender difference for any racial/ethnic group we note that for the 14 independent measurements shown in the figure there is higher than 30\% probability that one of the measurements will be found as far from zero as one sees for group AF and so we conclude that there is \textbf{no reliable evidence of a gender effect (even for group AF)}.

\begin{table}[htb]
\caption{The racial/ethnic groups in Fig. \ref{fig:GndrGapByRaceCourseA}, as described in the data obtained from the administration of the R1 university of this paper.}
\label{tab:tabA1}
\begin{ruledtabular}
\begin{tabular}{c c}
\textbf{Symbol} & \textbf{Ethnicity} \\ 
\hline
AF & African-American/Black\\
CH & Chinese-American/Chinese\\
EI & East Indian/Pakistani \\
FP & Filipino/Filipino-American \\
JA  & Japanese-American/Japanese 
\\
KO & Korean-American/Korean \\
LA & Latino/Other Spanish \\
MX & Mexican-American/Mexican/ \\
& Chicano\\
NA & Indigenous American/American Indian/ \\
& Native American \\
OA & Other Asian-American/Other Asian \\
VT & Vietnamese-American/Vietnamese \\
WH & White/Caucasian \\
\end{tabular}
\end{ruledtabular}
\end{table}

\section{\label{sec:CourseB}CourseII results for comparison}

In section \ref{sec.FracAsCorrelations} we showed that, for CourseI, the fraction of student answers that received an A was strongly positively correlated with course grade, strongly negatively correlated with the student's fraction of D and/or Non-zero F grades, but much more weakly correlated with the student's fraction of answers left blank.  In Table \ref{tab:BCorrelations} we show that the same is true of the grade data from CourseII.

\begin{table}[htbp]
\caption{The correlation coefficients between the fraction of exam-item A-grades given to a student in CourseII and three other student-level grade variables.  The fraction of A-grades is, as expected, strongly correlated with the course grade and the fraction of low grades (Ds or Fs) but has a much weaker correlation with the fraction of answers left blank.}
\label{tab:BCorrelations}
\begin{ruledtabular}
\begin{tabular}{c c c}
\textbf{Grade} & \textbf{Correlation between} &\textbf{P-value}\\ 
\textbf{variable} & \textbf{variable and FracAs} &\\ 
\hline
$CourseGrade$ & 0.806 & $<10^{-4}$ \\
$FracDsOrFs$ & -0.726 & $<10^{-4}$ \\
$FracBlanks$ & -0.315 & $<10^{-4}$ \\
\end{tabular}
\end{ruledtabular}
\end{table}

In Sec. \ref{sec:OverallDemographics} and Appendix \ref{sec:AllGenData} we showed that in CourseI classes there was no obvious gender difference in the choice to leave an answer blank.  We took this to be more evidence that the decision to leave a question unanswered is not simply related to not being able to answer the question.  In Table \ref{tab:BAllGroupGendGaps} we show that in CourseII there is also no gender gap in blank-leaving.  The two courses cover different topics in introductory physics so it seems that the topics covered do not strongly influence the decisions to leave answers blank and that the results are both qualitatively and quantitatively similar in the two courses.

\begin{table}[htbp]
\caption{The CourseII calculations for gender gap in fraction of blank answers, to compare with the CourseI calculations of Table \ref{tab:AllGroupGendGaps}.  Fraction of answers left blank in CourseII still seems independent of gender for the demographic groups.}
\label{tab:BAllGroupGendGaps}
\begin{ruledtabular}
\begin{tabular}{c c c c c c}
\textbf{HM} & \textbf{1stGen} & \textbf{GendGap} & \textbf{Error} & \textbf{P-value} & \textbf{N} \\
\hline
- & - & 0.00083 & 0.00055 & 0.132 & 12,200\\
0 & - & 0.00071 & 0.00058 & 0.220 & 10,453 \\
1 & - & 0.0020 & 0.0019 & 0.290 & 1,491 \\
- & 0 & 0.00091 & 0.00077 & 0.237 & 4,880 \\
- & 1 & 0.0014 & 0.0013 & 0.254 & 2,566 \\
0 & 0 & 0.00118 & 0.00079 & 0.134 & 4,336 \\
0 & 1 & 0.0007 & 0.0013 & 0.615 & 2,085 \\
1 & 0 & 0.0013 & 0.0034 & 0.693 & 424 \\
1 & 1 & 0.0040 & 0.0040 & 0.311 & 443 \\
\end{tabular}
\end{ruledtabular}
\end{table}

\begin{figure} [htb]
\includegraphics[trim=2.2cm 2.8cm 3.3cm 3.7cm, clip=true,width=\linewidth]{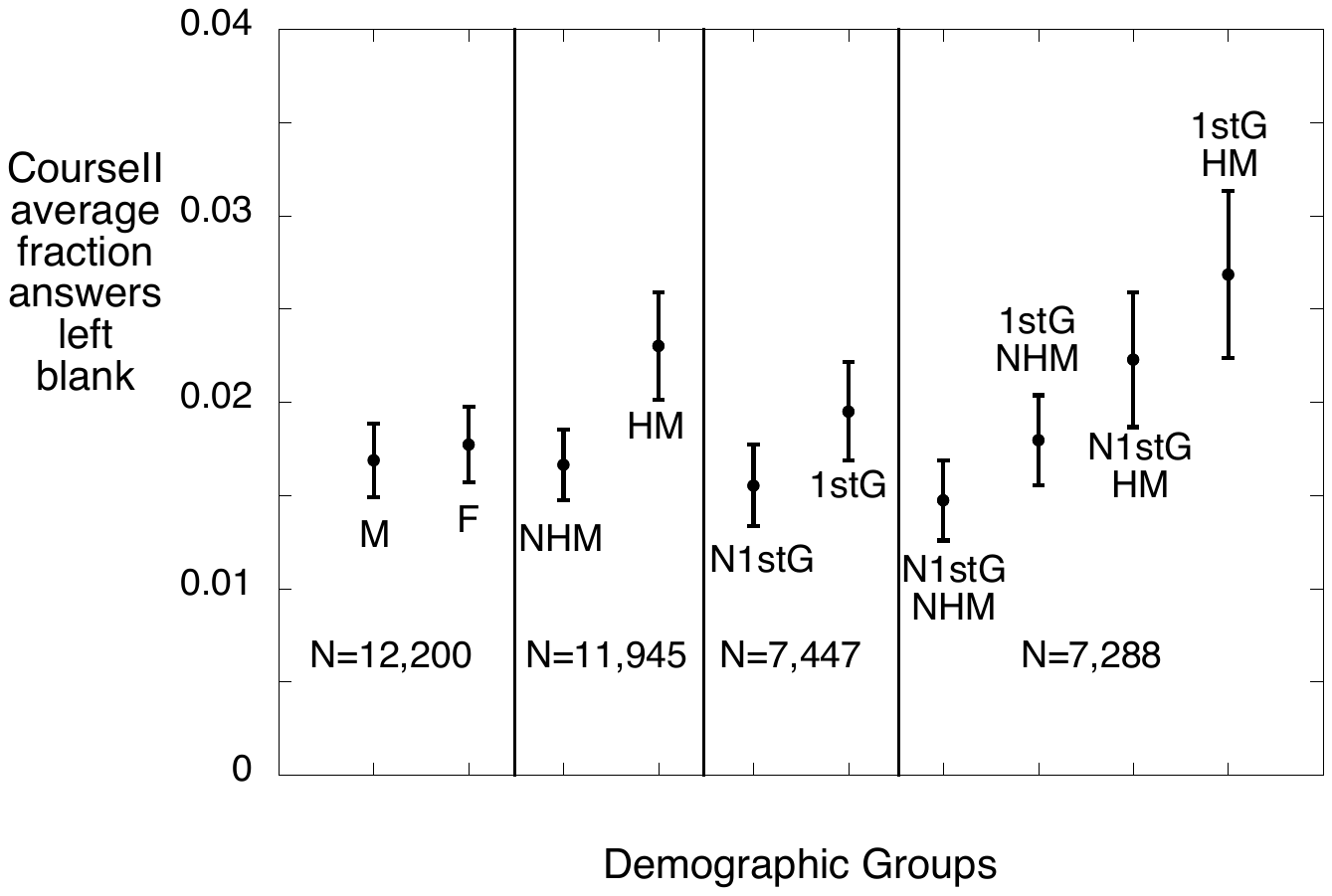}
\caption{The fraction of answers left blank, in the second term of the course series, is shown for the same demographic groups as Fig. \ref{fig:FracBlanksInAAllDemographics}.}
\label{fig:FracBlanksInBAllDemographics}
\end{figure}

Despite the fact that there are no obvious gender differences regarding leaving blanks, in Sec. \ref{sec:OverallDemographics} we showed that there are significant blank-leaving differences in between the demographic groups made up of the intersections of HM grouping and 1stGen grouping in the CourseI classes.  In Fig. \ref{fig:FracBlanksInBAllDemographics} we show that we find the same results when we include all students from CourseII classes.  Again, it seems that the topics covered do not strongly influence the decisions to leave answers blank and that the results are both qualitatively and quantitatively similar in the two courses.

\begin{figure} [htb]
\includegraphics[trim=2.2cm 3.0cm 2.6cm 4.0cm, clip=true,width=\linewidth]{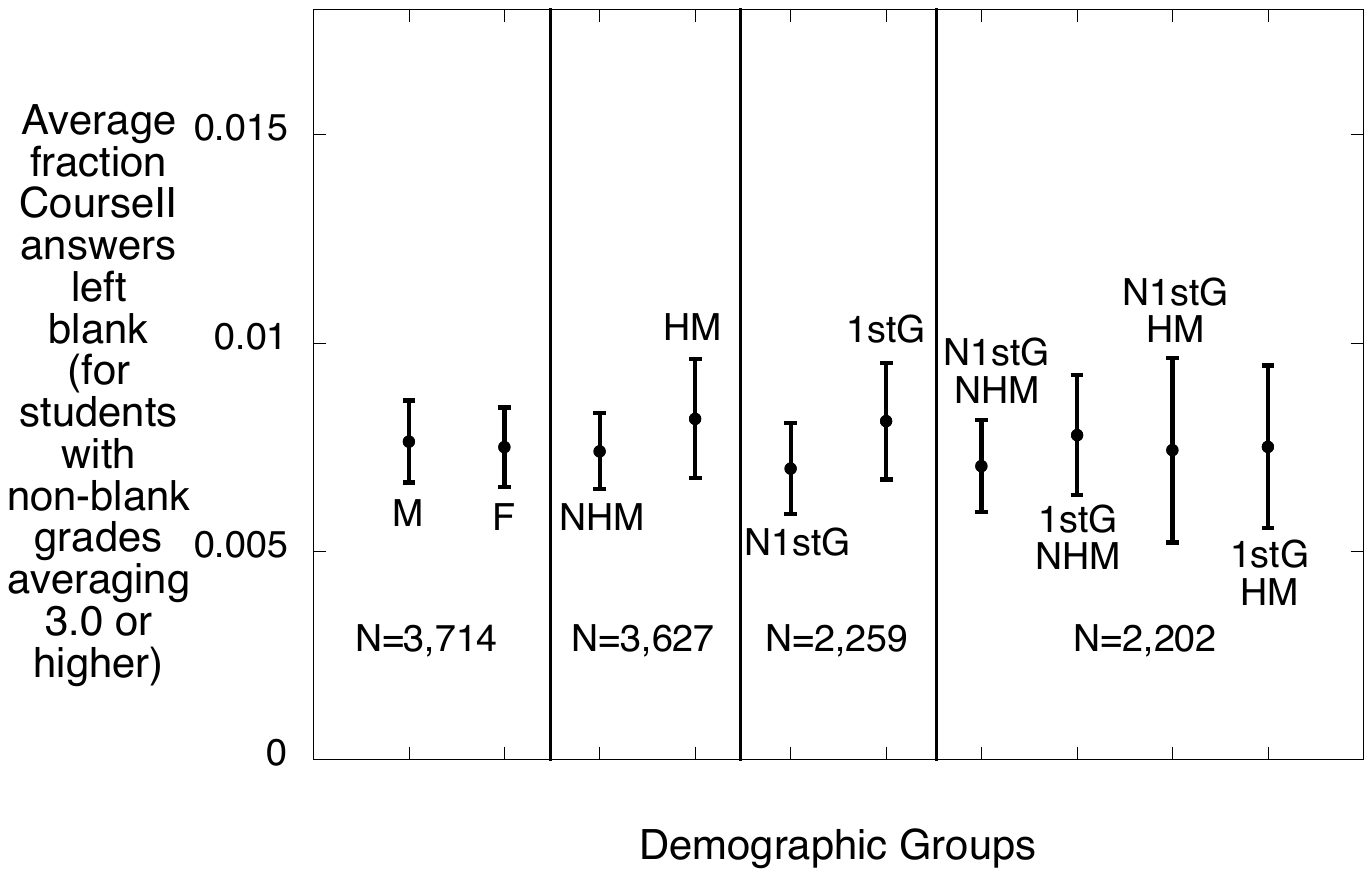}
\caption{The fraction of answers left blank, in the second term of the course series, is shown for the same demographic groups and for the same high scoring students (average non-blank grade is 3.0 or higher) as Fig. \ref{fig:FracBlanksInABGradeDemographics}.}
\label{fig:FracBlanksInBBGradesAllDemographics}
\end{figure}

In Fig. \ref{fig:FracBlanksInABGradeDemographics} we showed that the limiting our analyses to CourseI students who were doing well in their class (non-blank grade of middle-B or higher) did not change our conclusion that different demographic groups left different numbers of blanks on average.  For CourseII the analogous plot is shown in Fig.\ref{fig:FracBlanksInBBGradesAllDemographics}.  We may be seeing the first subtle difference between CourseI and CourseII blank-leaving.  In CourseII blank-leaving among the highest performing students may have been more independent of demographic group than was found for CourseI.  Of course there are some selection effects in moving from CourseI data to CourseII data that seem likely to affect the results.  Specifically, about 9\% of students from CourseI who were neither HM nor 1stGen did not move on to take CourseII.  In comparison, about 14\% of students in CourseI who had \textbf{both} HM and 1stGen status did not move on to take CourseII.  Similarly, students who left blanks in CourseI are about 50\% more likely to skip CourseII.  So, the differences between Fig. \ref{fig:FracBlanksInBBGradesAllDemographics} and Fig. \ref{fig:FracBlanksInABGradeDemographics} are likely influenced by selection effects changing different demographic groups differently.  In addition, there could also be some sort of training effects of CourseI on student behavior.  These things make any direct comparisons difficult.

\section{\label{sec:AllEthnicites}Blank-leaving choices by Racial/Ethnic groups}

\begin{figure} [htb]
\includegraphics[trim=2.3cm 3.1cm 3.0cm 3.9cm, clip=true,width=\linewidth]{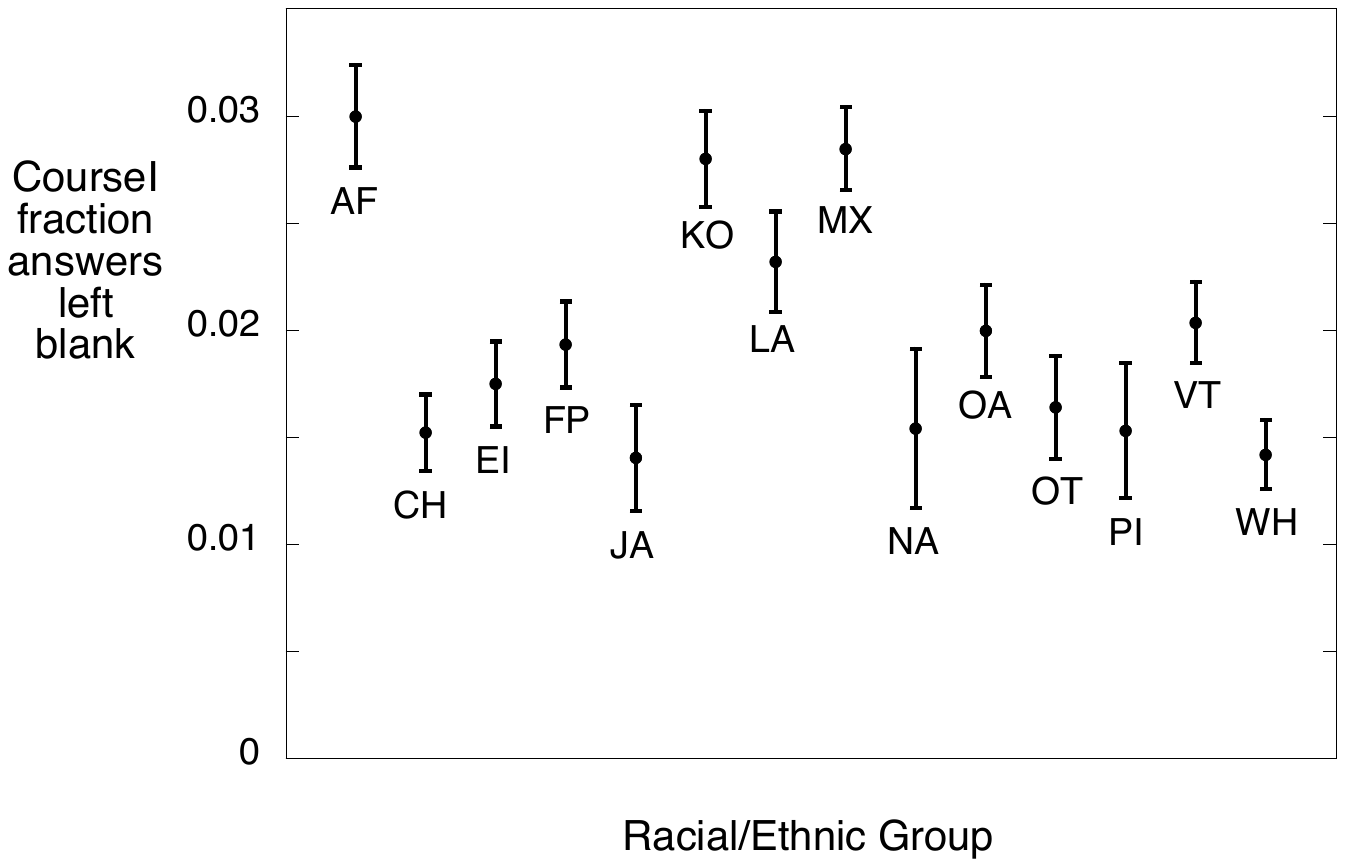}
\caption{The average fraction of CourseI answers left blank is shown for each of the racial/ethnic groups defined in Table \ref{tab:tabA1}}.
\label{fig:AFrc0Ethn}
\end{figure}

For completeness, we will show the fraction of answers left blank broken down by ethnicity.  We again use HLM to compute the average fraction of answers left blank in CourseI classes for each racial/ethnic group shown in Table \ref{tab:tabA1}.  Figure \ref{fig:AFrc0Ethn} shows these averages.  As noted in the main body of the paper we see that the three groups (AF, LA, and MX) that make up about 89\% of the HM demographic group left more blanks than their peers and we noted that Korean-American students also left more blanks than their peers.

\bibliography{Blanks}

@MISC{APS, 
author = {{American Physical Society}},
title = {{Underrepresented Minorities in Physics}}, 
howpublished = {website: aps.org/programs/education/statistics/urm.cfm}, 
month = {}, 
year = {}, 
note = {},
}

@article{Cotner2017,
author = {Cotner, Sehoya and Ballen, Cissy J.},
issn = {19326203},
journal = {PLoS ONE},
month = {dec},
number = {12},
pmid = {29281676},
publisher = {Public Library of Science},
title = {{Can mixed assessment methods make biology classes more equitable?}},
volume = {12},
year = {2017},
pages = {e0189610},
url = {https://doi.org/10.1371/journal.pone.0189610}
}

@article{Rodriguez2012,
author = {Rodriguez, Idaykis and Brewe, Eric and Sawtelle, Vashti and Kramer, Laird H},
doi = {10.1103/physrevstper.8.020103},
issn = {1554-9178},
journal = {Phys. Rev. ST Phys. Educ. Res.},
month = {jul},
number = {2},
pages = {020103+},
title = {{Impact of equity models and statistical measures on interpretations of educational reform}},
url = {http://dx.doi.org/10.1103/physrevstper.8.020103},
volume = {8},
year = {2012}
}

@article{Paul2022,
  title = {Percent grade scale amplifies racial or ethnic inequities in introductory physics},
  author = {Paul, Cassandra A. and Webb, David J.},
  journal = {Phys. Rev. Phys. Educ. Res.},
  volume = {18},
  issue = {2},
  pages = {020103},
  numpages = {16},
  year = {2022},
  month = {Jul},
  publisher = {American Physical Society},
  doi = {10.1103/PhysRevPhysEducRes.18.020103},
  url = {https://doi.org/10.1103/PhysRevPhysEducRes.18.020103}
}

@article{Webb2017,
author = {Webb,D. J. },
title = {Concepts first: A course with improved educational outcomes and parity for underrepresented minority groups},
journal = {Am. J. Phys.},
volume = {85},
number = {8},
pages = {628-632},
year = {2017},
doi = {10.1119/1.4991371},
URL = {https://doi.org/10.1119/1.4991371},
eprint = {https://doi.org/10.1119/1.4991371}
}

@article{Potter2014Sixteen,
author = {Potter, Wendell and Webb, David and Paul, Cassandra and West, Emily and Bowen, Mark and Weiss, Brenda and Coleman, Lawrence and {De Leone}, Charles},
doi = {10.1119/1.4857435},
issn = {0002-9505},
journal = {Am. J. Phys.},
month = {feb},
number = {2},
pages = {153--163},
title = {{Sixteen years of collaborative learning through active sense-making in physics (CLASP) at UC Davis}},
url = {http://dx.doi.org/10.1119/1.4857435},
volume = {82},
year = {2014}
}

@article{Webb2020,
  title = {Relative impacts of different grade scales on student success in introductory physics},
  author = {Webb, David J. and Paul, Cassandra A. and Chessey, Mary K.},
  journal = {Phys. Rev. Phys. Educ. Res.},
  volume = {16},
  issue = {2},
  pages = {020114},
  numpages = {17},
  year = {2020},
  month = {Aug},
  doi = {10.1103/PhysRevPhysEducRes.16.020114},
  url = {https://link.aps.org/doi/10.1103/PhysRevPhysEducRes.16.020114}
}

@article{Paul2014,
author = {Paul,Cassandra  and Potter,Wendell  and Weiss,Brenda },
title = {Grading by Response Category: A simple method for providing students with meaningful feedback on exams in large courses},
journal = {The Phys. Teach.},
volume = {52},
number = {8},
pages = {485-488},
year = {2014},
doi = {10.1119/1.4897587},
}

@article{Simmons2020,
	title = {Grades, grade component weighting, and demographic disparities in introductory physics},
	volume = {16},
	issn = {24699896},
	url = {https://doi.org/10.1103/PhysRevPhysEducRes.16.020125},
	doi = {10.1103/PhysRevPhysEducRes.16.020125},
	number = {2},
	journal = {Phys. Rev. Phys. Educ. Res.},
	author = {Simmons, Amber B. and Heckler, Andrew F.},
	year = {2020},
	pages = {020125},
}

@article{Paul2017,
	title = {Equity of success in {CLASP} courses at {UC} {Davis}},
	url = {https://www.compadre.org/per/items/detail.cfm?ID=14628},
	doi = {10.1119/perc.2017.pr.068},
	journal = {2017 Physics Education Research Conference Proceedings},
	author = {Paul, Cassandra A. and Webb, David J. and Chessey, Mary K. and Potter, Wendell H.},
	year = {2017},
	pages = {292--295},
}

@article{webb_attributing_2023,
	title = {Attributing equity gaps to course structure in introductory physics},
	volume = {19},
	url = {https://link.aps.org/doi/10.1103/PhysRevPhysEducRes.19.020126},
	doi = {10.1103/PhysRevPhysEducRes.19.020126},
	number = {2},
	urldate = {2023-10-17},
	journal = {Phys. Rev. Phys. Educ. Res.},
	author = {Webb, David J. and Paul, Cassandra A.},
	month = sep,
	year = {2023},
	pages = {020126},
}

@inproceedings{Paul2018,
author = {Paul, Cassandra and Webb, David J and Chessey, Mary K and Lucas, James},
booktitle = {2018 PERC Proceedings},
doi = {10.1119/perc.2018.pr.Paul},
url = {https://doi.org/10.1119/perc.2018.pr.Paul},
editor = {Traxler, A and Cao, Y and Wolf, S},
title = {{Pondering zeros : Uncovering hidden inequities within a decade of grades}},
year = {2018}
}

@article{paul_examining_2025,
    title = {Examining equity and graduation rates at two institutions using a course deficit model and the collaborative learning through active sense-making in physics curriculum},
    volume = {21},
    issn = {2469-9896},
    url = {https://doi.org/10.1103/PhysRevPhysEducRes.21.010107},
    doi = {10.1103/PhysRevPhysEducRes.21.010107},
    number = {1},
    urldate = {2025-03-06},
    journal = {Phys. Rev. Phys. Educ. Res.},
    author = {Paul, Cassandra A. and Webb, David J.},
    month = jan,
    year = {2025},
    pages = {010107}
}

@article{stephens_unseen_2012,
	title = {Unseen disadvantage: {How} {American} universities' focus on independence undermines the academic performance of first-generation college students.},
	volume = {102},
	issn = {1939-1315, 0022-3514},
	shorttitle = {Unseen disadvantage},
	url = {http://doi.apa.org/getdoi.cfm?doi=10.1037/a0027143},
	doi = {10.1037/a0027143},
	number = {6},
	urldate = {2025-07-21},
	journal = {Journal of Personality and Social Psychology},
	author = {Stephens, Nicole M. and Fryberg, Stephanie A. and Markus, Hazel Rose and Johnson, Camille S. and Covarrubias, Rebecca},
	year = {2012},
	note = {Publisher: American Psychological Association (APA)},
	pages = {1178--1197},
}

@article{stathopoulou_language_2006,
	title = {Language and {Culture} in {Mathematics} {Education}: {Reflections} on {Observing} a {Romany} {Class} in a {Greek} {School}},
	volume = {64},
	copyright = {http://www.springer.com/tdm},
	issn = {0013-1954, 1573-0816},
	shorttitle = {Language and {Culture} in {Mathematics} {Education}},
	url = {http://link.springer.com/10.1007/s10649-006-4496-y},
	doi = {10.1007/s10649-006-4496-y},
	number = {2},
	urldate = {2025-07-21},
	journal = {Educational Studies in Mathematics},
	author = {Stathopoulou, Charoula and Kalabasis, Fragiskos},
	month = dec,
	year = {2006},
	note = {Publisher: Springer Science and Business Media LLC},
	pages = {231--238},
}

@article{decaro_active_2025,
	title = {Active {Learning} {Increases} {Gateway} {Engineering} {Course} {Outcomes} and {Closes} {Preparation} {Gap} for {Underrepresented} {Students}},
	issn = {2520-8705, 2520-8713},
	url = {https://link.springer.com/10.1007/s41979-025-00159-y},
	doi = {10.1007/s41979-025-00159-y},
	urldate = {2025-10-08},
	journal = {Journal for STEM Education Research},
	author = {DeCaro, Marci S. and Golway, Katherine G. and Bego, Campbell R. and Derkson, Shannon N. and Hieb, Jeffrey L.},
	month = sep,
	year = {2025},
}

@article{webb_highstakes_2025,
	title = {High-stakes exams inflate a gender gap and contribute to systematic grading errors in one introductory physics series},
	volume = {21},
	issn = {2469-9896},
	url = {https://link.aps.org/doi/10.1103/ngrs-dxtf},
	doi = {10.1103/ngrs-dxtf},
	number = {2},
	urldate = {2025-10-22},
	journal = {Physical Review Physics Education Research},
	author = {Webb, David J. and Paul, Cassandra A.},
	month = oct,
	year = {2025},
	pages = {020134},
}

@misc{young_reducing_2025,
	title = {Reducing the weight of low exam scores may raise average grades but does not appear to impact equity gaps},
	url = {http://arxiv.org/abs/2508.09906},
	doi = {10.48550/arXiv.2508.09906},
	urldate = {2026-02-27},
	publisher = {arXiv},
	author = {Young, Nicholas T. and Matz, Rebecca L. and Bell, Eric F. and Hayward, Caitlin},
	month = aug,
	year = {2025},
	note = {arXiv:2508.09906 [physics]},
}

@article{Reeves2004,
    title = {The {Case} {Against} the {Zero}},
    volume = {86},
    url = {https://doi.org/10.1177/003172170408600418},
    number = {4},
    journal = {Phi Delta Kappan},
    author = {Reeves, Douglas B.},
    year = {2004},
    pages = {324--325},
}

@article{Guskey2013,
    title = {The case against percentage grades},
    volume = {71},
    doi = {10.1136/bmj.314.7085.970},
    number = {1},
    journal = {Educ. Leadersh.},
    author = {Guskey, T R},
    year = {2013},
    pages = {68--72},
}

@article{Beatty2013,
    title = {Standards-based grading in introductory university physics},
    volume = {13},
    number = {2},
    journal = {J. Scholarsh. Teach. Learn.},
    author = {Beatty, I D},
    year = {2013},
    pages = {1--22},
}

@phdthesis{Chessey2018,
author={Chessey,Mary K.},
school = {University of Calif., Davis},
year={2018},
title={Transfer Students Redefining Physics Culture: Student Agency and Responses to Traditional Physics Education},
journal={ProQuest Dissertations and Theses},
pages={111},
note={Copyright - Database copyright ProQuest LLC; ProQuest does not claim copyright in the individual underlying works; Last updated - 2023-03-03},
isbn={978-0-438-29144-7},
url={https://www.proquest.com/dissertations-theses/transfer-students-redefining-physics-culture/docview/2095900296/se-2},
}

@article{Townsley2020,
	title = {Alternative grading practices: {An} entry point for faculty in competency‐based education},
	volume = {5},
	issn = {2379-6154},
	doi = {10.1002/cbe2.1219},
	number = {3},
	journal = {The Journal of Competency-Based Education},
	author = {Townsley, Matt and Schmid, David},
	year = {2020},
	pages = {1--5},
}

@article{hackerson_alternative_2024,
	title = {Alternative grading practices in undergraduate {STEM} education: a scoping review},
	volume = {6},
	issn = {2662-2300},
	shorttitle = {Alternative grading practices in undergraduate {STEM} education},
	url = {https://doi.org/10.1186/s43031-024-00106-8},
	doi = {10.1186/s43031-024-00106-8},
	number = {1},
	urldate = {2024-05-06},
	journal = {Disciplinary and Interdisciplinary Science Education Research},
	author = {Hackerson, Emily L. and Slominski, Tara and Johnson, Nekeisha and Buncher, John B. and Ismael, Safana and Singelmann, Lauren and Leontyev, Alexey and Knopps, Alexander G. and McDarby, Ariana and Nguyen, Jonathan J. and Condry, Danielle L. J. and Nyachwaya, James M. and Wissman, Kathryn T. and Falkner, William and Grieger, Krystal and Montplaisir, Lisa and Hodgson, Angela and Momsen, Jennifer L.},
	month = may,
	year = {2024},
	pages = {15},
}

@article{leslie_specifications_2020,
	title = {Specifications {Grading}: {What} {It} {Is}, and {Lessons} {Learned}},
	volume = {41},
	issn = {0734-0478, 1098-9056},
	shorttitle = {Specifications {Grading}},
	url = {http://www.thieme-connect.de/DOI/DOI?10.1055/s-0040-1713781},
	doi = {10.1055/s-0040-1713781},
	
	number = {04},
	urldate = {2026-03-20},
	journal = {Seminars in Speech and Language},
	author = {Leslie, Paula and Lundblom, Erin},
	month = aug,
	year = {2020},
	pages = {298--309},
}

@inproceedings{armacost_using_2003,
	address = {Westminster, Colorado, USA},
	title = {Using mastery-based grading to facilitate learning},
	volume = {1},
	isbn = {978-0-7803-7961-9},
	url = {http://ieeexplore.ieee.org/document/1263320/},
	doi = {10.1109/FIE.2003.1263320},
	
	urldate = {2026-03-20},
	booktitle = {33rd {Annual} {Frontiers} in {Education}, 2003. {FIE} 2003.},
	publisher = {IEEE},
	author = {Armacost, R.L. and Pet-Armacost, J.},
	year = {2003},
	pages = {T3A\_20--T3A\_25},
}

@article{speirs_thematic_2026,
	title = {Thematic analysis of students’ perceptions of grading practices in physics},
	volume = {22},
	issn = {2469-9896},
	url = {https://link.aps.org/doi/10.1103/pyvm-4s9t},
	doi = {10.1103/pyvm-4s9t},
	 
	number = {1},
	urldate = {2026-03-20},
	journal = {Physical Review Physics Education Research},
	author = {Speirs, J. Caleb and Lane, W. Brian and Laird, Naomi},
	month = feb,
	year = {2026},
	pages = {010115},
}

@article{Carey2012,
author = {Carey, T. and Carifio, J.},
doi = {10.3102/0013189X12453309},
journal = {Educ. Res.},
number = {6},
pages = {201--208},
title = {{The Minimum Grading Controversy: Results of a Quantitative Study of Seven Years of Grading Data From an Urban High School}},
url = {http://edr.sagepub.com/cgi/doi/10.3102/0013189X12453309},
volume = {41},
year = {2012}
}

@article{mattersMultipleChoiceShortResponseItems1999,
  title = {Multiple-{{Choice}} versus {{Short-Response Items}}: {{Differences}} in {{Omit Behaviour}}},
  shorttitle = {Multiple-{{Choice}} versus {{Short-Response Items}}},
  author = {Matters, Gabrielle and Burnett, Paul C.},
  year = 1999,
  month = aug,
  journal = {Australian Journal of Education},
  volume = {43},
  number = {2},
  pages = {117--128},
  issn = {0004-9441, 2050-5884},
  doi = {10.1177/000494419904300202},
  urldate = {2026-09-04},
  copyright = {https://journals.sagepub.com/page/policies/text-and-data-mining-license},
  langid = {english}
}

@article{dichiacchioExaminingHowMotivation2016,
  title = {Examining How Motivation toward Science Contributes to Omitting Behaviours in the {{Italian PISA}} 2006 Sample},
  author = {Di Chiacchio, Carlo and De Stasio, Simona and Fiorilli, Caterina},
  year = 2016,
  month = aug,
  journal = {Learning and Individual Differences},
  volume = {50},
  pages = {56--63},
  issn = {10416080},
  doi = {10.1016/j.lindif.2016.06.025},
  urldate = {2026-09-04},
  langid = {english}
}

\end{document}